\documentclass[twocolumn,pre, 10pt, floatfix]{revtex4-1}

\usepackage{graphicx, amsmath, amsfonts, amssymb, xcolor}

\usepackage[utf8]{inputenc}

\usepackage[english]{babel}

\newcommand{\ud}{\mathrm{d}}

\graphicspath{{figures/}}
\begin{document}

\title{Analytical approximations for the speed of pacemaker-generated waves}

\author{Jan Rombouts}
\author{Lendert Gelens}
\affiliation{Laboratory of Dynamics in Biological Systems, Department of Cellular 
and Molecular Medicine, University of Leuven (KU Leuven), B-3000 Leuven, Belgium}

\begin{abstract}
In an oscillatory medium, a region which oscillates faster than its surroundings can act as a source of outgoing waves. Such
pacemaker-generated waves can synchronize the whole medium and are present in many chemical and biological systems, where they are a means of transmitting information at a fixed speed over large distances. In this paper, we apply analytical tools to investigate the factors that determine the speed of these waves. More precisely, we apply singular perturbation and phase reduction methods to two types of negative-feedback oscillators, one built on underlying bistability and one including a time delay in the negative feedback. In both systems, we investigate the influence of timescale separation on the resulting wave speed, as well as the effect of size and frequency of the pacemaker region. We compare our analytical estimates to numerical simulations which we described previously \cite{rombouts_synchronizing_2020}. 
\end{abstract}

\maketitle

\section{Introduction}

Many different biological and chemical systems show rhythmic behavior, characterized by the periodic rise and fall of activities and concentrations of chemicals. This periodic behavior is typically the result of an intricate network of chemical reactions, where proteins or other chemicals interact through multiple nonlinear feedback loops. One necessary condition for such a system to oscillate is the presence of a negative feedback loop, which is needed to reset the system after a cycle. In order to ensure robust, large-amplitude oscillations, the negative feedback is complemented by additional mechanisms in many systems. Two of the main mechanisms that enhance robust oscillatory behavior are a time delay or a positive feedback loop \cite{novak_design_2008}.

If an oscillatory system is extended in space and molecules diffuse, different types of spatiotemporal patterns can form such as target patterns, spiral waves or chaos. Traveling waves and target patterns are often generated by pacemakers: regions in the medium that oscillate at a higher frequency than the surroundings. Classic examples of such waves are found in the Belousov-Zhabotinsky reaction and its variants \cite{taylor_mechanism_2002} or the oxidation of CO on Pt(110) \cite{jakubith_spatiotemporal_1990}. In these chemical systems, a pacemaker is often the result of a particle of dust or a small deformation in the reaction dish, but it can also be induced, for example using a laser \cite{wolff_laser-induced_2004}. 
In biology, such pacemaker-generated waves can be seen in cAMP signaling in populations of \textit{Dictyostelium discoideum}  \cite{bretschneider_progress_2016} or in the spreading of cell cycle events through extracts made of \textit{Xenopus laevis} eggs \cite{chang_mitotic_2013, nolet_nuclei_2020, afanzar_nucleus_2020}. Traveling waves also appear in electrical, rather than biochemical, systems: in neural and cardiac tissue, for example. Recent overviews of biochemical oscillations and waves with many examples can be found in Refs. \cite{beta_intracellular_2017,deneke_chemical_2018}.

In large biological systems, these waves can be a means of transmitting information over large distances as they propagate at a constant speed, whereas diffusion by itself is too slow \cite{gelens_spatial_2014}. In such systems, it is important to know what determines the speed of these waves. A host of different factors influence the wave speed: the dynamics of the oscillator itself, the diffusion of the surrounding medium and the properties of the pacemaker region such as its size and frequency. 

In a previous paper \cite{rombouts_synchronizing_2020}, we described our numerical results on the speed of pacemaker-generated waves. There, we demonstrated the difference between an oscillator built on an underlying bistability and one in which the oscillations were generated by a time-delayed negative feedback loop. Whereas the time series of these systems are very similar, the waves they produce show marked differences. Our numerical results were based on explicit simulations of the partial differential equations of the reaction-diffusion system. 

In this paper, we take a complementary approach based on analytical methods that have been traditionally used to analyze wave propagation. There are numerous mathematical studies on waves, patterns and pacemakers in oscillatory media. However, those studies typically focus on a limiting case. For systems close to a Hopf bifurcation, the dynamics of a spatially extended system can be described by the complex Ginzburg-Landau equation (CGLE) \cite{aranson_world_2002, garcia-morales_complex_2012}. Many studies take this equation as a starting point, since it is analytically tractable. \citet{stich_complex_2002} analyzed waves generated by a pacemaker in the CGLE in one dimension. The same authors also analyzed the CGLE in two dimensions both for circular and rectangular inhomogeneities \cite{stich_target_2006}. 
Related to the CGLE are $\lambda-\omega$ systems, which also show harmonic oscillations. For these systems, mathematically rigorous results on the existence of target patterns have been obtained \cite{kopell_target_1981}. 

If the influence of diffusion can be considered small, a powerful technique called the phase reduction method can be used, in which one makes the assumption that diffusion only affects the phase of the oscillator on its limit cycle, but not its amplitude. Such an approach has been used by \citet{neu_chemical_1979}, \citet{hagan_target_1981} and Kuramoto \cite{kuramoto_chemical_1984} to obtain qualitative insight into the existence and characteristics of waves in oscillatory media. These authors derive the phase equation starting from the underlying reaction-diffusion system. However, as with the CGLE, often the phase equation is taken as the starting point of an analysis. 

A different method to describe waves is the singular perturbation technique. The method is usually applied to excitable systems. In these systems, small perturbations to a steady state die out, but large perturbations trigger a large response before returning to the steady state. Such excitable systems support traveling waves when they are extended in space. The singular perturbation method can also be used to describe oscillatory media and can be applied when the underlying dynamics show strong timescale separation. In that limit, the waves in such systems can be treated as traveling fronts, whose speed can be determined. This method is described in many papers \cite{tyson_singular_1988,meron_pattern_1992,casten_perturbation_1975,keener_geometrical_1986} and textbooks (e.g. \cite[Ch. 6]{keener_mathematical_2009}).

Both methods --- phase dynamics and singular perturbation --- have led to a lot of insight into the qualitative dynamics of traveling waves and spatial patterns. However, they are rarely used to obtain a concrete estimate of the wave speed in a given reaction-diffusion system. In this paper, our aim is to explain how these methods can be used to obtain an estimate of the wave speed. We do this for two types of oscillator, and focus on the influence of a timescale separation parameter and on features of the pacemaker itself. We compare the analytical estimates to our numerical results reported previously. In Section~\ref{sec:setup}, we describe the equations we use and define the spatial setup of the system. We explain how the singular perturbation method can be used to obtain an approximation in the limit of large timescale separation in Section~\ref{sec:singperturb}. The phase reduction method is explained in Section~\ref{sec:phasered}, and here we focus on the properties of the pacemaker itself such as its size and frequency. We end with a discussion in Section~\ref{sec:discussion}. 

\section{Setup} \label{sec:setup}

We will consider two different mechanisms to generate oscillations, which nevertheless produce very similar time series. 

\begin{figure}
 \includegraphics{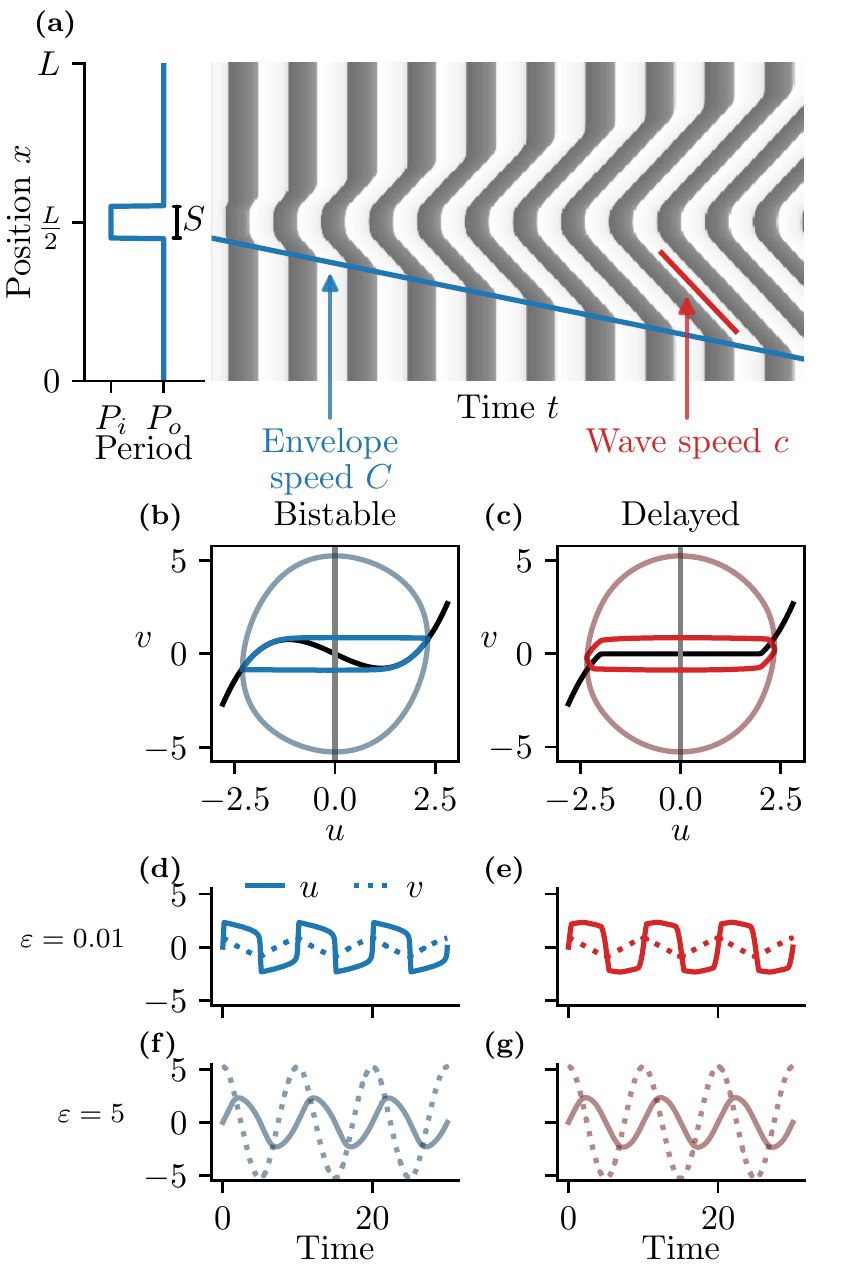}
 \caption{Setup of the system. (a) A pacemaker of size $S$ is situated in the middle of an oscillatory medium. The oscillation period of the medium is $P_o$, that of the pacemaker is $P_i$. The pacemaker sends out waves of speed $c$, which overtake the medium with an envelope speed $C$. The shade of gray corresponds to the value of $u$. (b,c) Limit cycles in the $(u,v)$-plane of bistable (b) and delayed oscillator (c). The cycle is shown for two values of the timescale separation parameter $\varepsilon$. (d-g) Time series of $u$ and $v$ for bistable (left) and delayed (right) oscillator. Small value of $\varepsilon$ in panels d,e and a larger value in panels f,g.}
 \label{fig:setup}
\end{figure}

The first model is based on the Van der Pol oscillator \cite{van_der_pol_relaxation-oscillations_1926} and the related FitzHugh-Nagumo equations \cite{fitzhugh_impulses_1961, nagumo_active_1962}. It reads
\begin{equation}
 \begin{aligned}
  u_t &= \varepsilon^{-1} (v - \frac{1}{4}u(u^2-4)) + D_u u_{xx} \\
  v_t &= -u + D_v v_{xx}. 
 \end{aligned}
  \label{eq:eqvdp}
\end{equation}
We will refer to this model as the \emph{bistable oscillator}, with slight abuse of terminology. The system as written down above does not exhibit bistability, only oscillations. However, the oscillations can be considered to be built on top of a bistable system: if $v$ is a constant, the equation for $u$ admits two stable solutions. This can be seen in the phase plane (Fig.~\ref{fig:setup}b). 

We will compare this system to the \emph{delayed oscillator} with equations
\begin{equation}
 \begin{aligned}
  u_t &= \varepsilon^{-1}(v(t-\tau) - \tilde f(u)) + D_u u_{xx} \\
  v_t &= -u + D_v v_{xx} \\
  \tilde f(u) &= f(u)\times \kappa_{[-2,2]}(u),
 \end{aligned}
 \label{eq:delayed}
\end{equation}
where $\kappa_{[-2,2]}(u) = 0$ if $|u| < 2$ and $1$ otherwise, and $f$ is the cubic polynomial used in Eq.~\eqref{eq:eqvdp}: $f(u) = \frac{1}{4} u (u^2-4)$. This system also admits limit-cycle solutions. The definition of $\tilde f$ was chosen in such a way that the limit cycle solution of the bistable and delayed oscillator are as similar as possible (Fig.~\ref{fig:setup}b--g). 

In this paper, we will use $D_u = D_v = 1$. A numerical study of the influence of the diffusion constants can be found in Ref.~\cite{rombouts_synchronizing_2020}. To modify the characteristics of the underlying oscillator, we focus on a single parameter: $\varepsilon$, which determines the timescale separation. This parameter mainly determines the nature of the oscillation (Fig.~\ref{fig:setup}b-g). If $\varepsilon$ is small, the oscillations show sharp jumps and are relaxation-like (Fig.~\ref{fig:setup}d,e). For larger values of $\varepsilon$, they become more sinusoidal (Fig.~\ref{fig:setup}f,g). The delayed oscillator has one extra parameter, namely the time delay. We link the delay $\tau$ to $\varepsilon$: for each $\varepsilon$ we consider, we fit $\tau$ such that the amplitude of the $v$-variable of the delayed oscillator is the same as that for the bistable oscillator.

Mathematically, these systems are different since one of them includes a time delay, whereas the other one is based on regular ODEs. This makes a comparison in the mathematical sense difficult. However, both systems show very similar time series. Conceptually, the time delay replaces the stable branches of the cubic nullcline, and produces the overshoot necessary for oscillations. If the only thing we know about a system would be its time series, we would not directly be able to say if the underlying mechanism that generates the oscillations is more like our bistable oscillator or more like the delayed one. From such a viewpoint, it is interesting to compare the properties of both systems when they are extended in space. 

In the spatial setup, we consider a one-dimensional domain of length $L$ with a pacemaker of size $S$ in the middle. The period of the pacemaker is $P_i$ and the period of the outside medium is $P_o$ (Fig.~\ref{fig:setup}a). These numbers are parameters of the system, which we impose in simulations by first computing the natural period of the system, and rescaling time in the simulation of the PDEs. 

The pacemaker sends out waves which overtake the medium. The waves themselves propagate at a speed $c$, the wave speed, and they overtake the medium with a speed we call the envelope speed, denoted by $C$ (Fig.~\ref{fig:setup}a). In our previous paper we showed that the envelope does not need to spread linearly \cite{rombouts_synchronizing_2020}. This can happen when $\varepsilon$ is large, or when the size of the pacemaker or its frequency difference with the outside medium is too small. In the current study, we only focus on the case where the envelope spreads linearly. In these cases, we can speak of a well-defined wave and envelope speed. 

We now explain how we can obtain an analytical estimate of the wave speed of the pacemaker-generated waves, first using a singular perturbation method and afterwards with the phase reduction approach.

\section{Singular perturbation} \label{sec:singperturb}

The singular perturbation method can be applied to the bistable oscillator for $\varepsilon \to 0$. We give a brief overview of the computation in our specific case. More general treatments can be found in Refs. \cite{tyson_singular_1988,meron_pattern_1992,keener_mathematical_2009}. 

We start from the equations
\begin{equation}
 \begin{aligned}
  u_t &= \varepsilon^{-1} K (v - \frac{1}{4}u(u^2-4)) + D_u u_{xx} \\
  v_t &= -Ku + D_v v_{xx}, 
 \end{aligned}
  \label{eq:eqvdpwithK}
\end{equation}
where we have added the parameter $K$. We do this because for each $\varepsilon$, we rescale time to obtain a period of $P_o$ in the medium. The value of $K$ thus depends on $\varepsilon$. However, it is of order 1 for low $\varepsilon$ and does not affect the computation. It needs to be included to obtain estimates that can be compared to the numerical results. 

We rescale time and space and set $s = Kt$ and $y = \sqrt{\varepsilon} x$. The equations can be written as 
\begin{equation}
 \begin{aligned}
  \varepsilon u_s &= v - \frac{1}{4}u(u^2-4) + \frac{D_u}{K} \varepsilon^2 u_{yy} \\
  v_s &= -u + \frac{D_v}{K}\varepsilon v_{yy}. 
 \end{aligned}
  \label{eq:eqvdprescaled}
\end{equation}

To look for traveling wave solutions, we set $z= y + cs$ and obtain
\begin{equation}
 \begin{aligned}
  \varepsilon c u_z &= v - \frac{1}{4}u(u^2-4) + \frac{D_u}{K} \varepsilon^2 u_{zz} \\
  c v_z &= -u + \frac{D_v}{K}\varepsilon v_{zz}. 
 \end{aligned}
  \label{eq:eqvdpz}
\end{equation}

\begin{figure}
 \centering
 \includegraphics{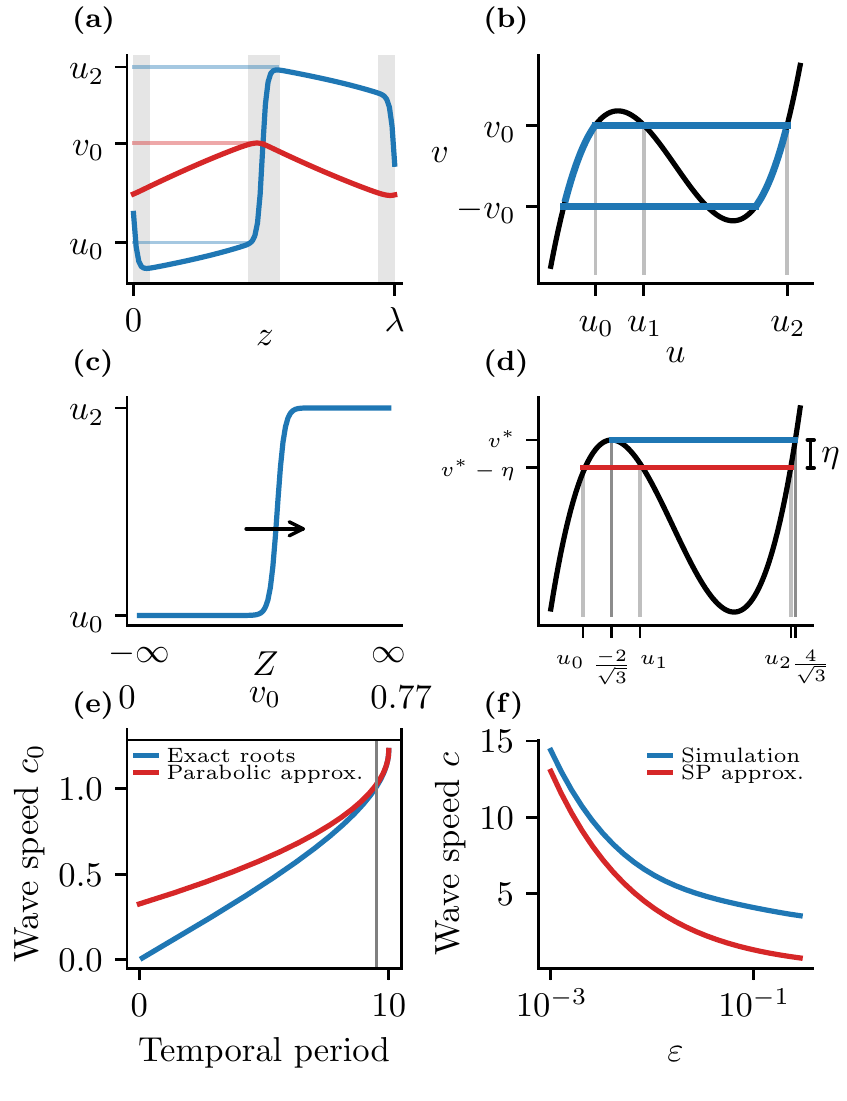}
 \caption{Singular perturbation. (a) The solution consists of slow progress along the branches and fast jumps in between: the outer and inner solutions respectively. Shaded rectangles indicated the inner solution. The blue curve denotes $u$, the red curve $v$. The value of $v$ at the jump is $v_0$.  (b) The value of $v_0$ determines the level of $v$ at the jump and the limits on the branches of the cubic nullcline that determine the period. (c) The inner solution is a bistable front. (d) Approximation of the roots of $v_0 - f(u)$ when $v_0$ is close to the value at the extremum of the cubic nullcline. The roots are small perturbations of $-2/\sqrt{3}$ and $4/\sqrt{3}$. (d) Dispersion relation obtained by numerically solving for the roots of $v_0 - f(u)$ in blue, and with the parabolic approximation in red. (f) Result for the wave speed in the original system, compared to the result from a numerical simulation. }
 \label{fig:singperturb}
\end{figure}

A wave train is a periodic solution of this system. The period in the $z$-variable corresponds to the spatial wavelength in the original system, which we call $\lambda$  (Fig.~\ref{fig:singperturb}a). This is the classical setup for the singular perturbation problem. For small $\varepsilon$, the solution consists of parts that evolve along the branches of the cubic in the phase plane (the outer solutions) and sharp jumps between those branches (inner solutions) (Fig.~\ref{fig:singperturb}a,b). 
First, we determine the outer solution to leading order. We set $\varepsilon = 0$ in Eq.~\eqref{eq:eqvdpz} to find
\begin{equation}
 \begin{aligned}
0 &= v - \frac{1}{4}u(u^2-4) \\
  c v_z &= -u. 
 \end{aligned}
\end{equation}
The first equation shows that the outer solution proceeds along the branches of the cubic nullcline. Taking the derivative with respect to $z$ of the first equation and substituting it into the second gives
\[ \frac{4-3u^2}{4u} u_z = \frac{1}{c}.\]
Integrating with respect to $z$ leads to
\[ \int \frac{4-3u^2}{4u} \ud u = \frac{1}{c} \int\ud z
\]
where the integral runs over a branch of the cubic nullcline for $u$ and one segment of the wave train for $z$. Because of the symmetry in our system, twice the integral over $z$ is the whole wavelength $\lambda$ to leading order. Since wavelength divided by speed gives the temporal period of the wave, twice the $u$-integral gives us the temporal period of the wave train. We can compute the value of this integral if we know its limits. These are determined by the value of $v$ at the jump between the two branches (Fig.~\ref{fig:singperturb}b). If the jump happens at $v=v_0$, we define $u_i$ to be the three intersection points of the line $v=v_0$ with the cubic function $f(u) = \frac{1}{4} u(u^2-4)$, as indicated in the figure. Using this terminology, and using the symmetry of the system, the integral for the right branch reads
\begin{equation}
\int_{u_2}^{-u_0} \frac{4-3u^2}{4u} \ud u = \ln\left|\frac{u_0}{u_2}\right| - \frac{3}{8} (u_0^2 - u_2^2). \label{eq:intu} 
\end{equation}

Next, we compute the inner solution. We use the stretched coordinate $Z=z/\varepsilon$. In this variable, the equations read
\begin{equation}
 \begin{aligned}
  c u_Z &= v - \frac{1}{4}u(u^2-4) + \frac{D_u}{K} u_{ZZ} \\
  c v_Z &= -\varepsilon u + \frac{D_v}{K}  v_{ZZ}. 
 \end{aligned}
  \label{eq:eqvdpZ}
\end{equation}

To leading order, this gives
\[
\begin{aligned}
  c u_Z &= v - \frac{1}{4}u(u^2-4) + \frac{D_u}{K} u_{ZZ} \\
  c v_Z &= \frac{D_v}{K} v_{ZZ}. 
 \end{aligned}
\]
The only bounded solution of the equation for $v$ is $v=v_0$, a constant. This gives us the following equation for $u$:
\begin{equation}
 \frac{D_u}{K} u_{ZZ} - cu_z + g(u)= 0, \label{eq:ufront}
\end{equation}
where $g$ is the cubic polynomial given by $g(u) = v_0 - \frac{1}{4}u(u^2-4)$. To match the outer solution, we impose conditions on the solution of this equation for $Z \to \pm \infty$. For the jump from the left to the right branch, the conditions are $u(Z) \to u_0$ for $Z \to -\infty$ and $u(Z) \to u_2$ for $Z\to\infty$ (Fig.~\ref{fig:singperturb}c). The values of $u_i$, which we defined above as the intersection points, are also the zeroes of the cubic function $g$. 

Eq.~\eqref{eq:ufront} is exactly the equation for a cubic bistable front. It admits a solution for a single value of $c$. In other words, given $v_0$, $c$ is fixed. Because $g$ is cubic, the solution can be computed analytically (e.g. Ref. \cite[Sec. 6.2]{keener_mathematical_2009}). In our case, this leads to 
\begin{equation}
 c = \sqrt{\frac{D_u}{8K}} (u_2 - 2u_1 + u_0). \label{eq:cfront}
\end{equation}

To summarize: for each value of $v_0$ we can find the roots $u_i$. The $u_i$ determine two things: the speed of a front connecting the two branches, and the limits of the integrals on the branches. This integral gives us the temporal period $T$. The spatial wavelength $\lambda$ is equal to $cT$. This procedure gives parametrizations of $c, T$ and $\lambda$ as function of $v_0$, which we can use to plot a dispersion relation for the wave train (Fig.~\ref{fig:singperturb}e). 

The values $u_i$ are roots of a third-order polynomial, which can in principle be computed analytically using the cubic formula. This would lead to long and messy equations. More insight can be obtained by computing an approximation of these roots, and of the speed and the period of the waves, when the jump between the branches happens near the extremum (Fig.~\ref{fig:singperturb}d). We consider the jump for $v>0$. The function $f$ has a maximum at $u = -2/\sqrt{3}$ with value $v^* = 4/(3\sqrt{3})$. Let us now consider $v = v^* - \eta$ with $\eta$ small. We can write down an asymptotic approximation for the roots:
\begin{equation}
 \begin{aligned}
  u_0 &\approx \frac{-2}{\sqrt{3}} - \sqrt{\frac{2}{\sqrt{3}}}\eta^{1/2} \\
  u_1 &\approx \frac{-2}{\sqrt{3}} + \sqrt{\frac{2}{\sqrt{3}}}\eta^{1/2} \\
  u_2 &\approx 4/\sqrt{3} - \frac{1}{3}\eta.
 \end{aligned}
\end{equation}

By substituting these into Eq.~\eqref{eq:cfront}, we find (keeping only terms up to $\eta^{1/2}$):

\[ c \approx \sqrt{\frac{D_u}{K}} \left(\sqrt{\frac{3}{2}} - \frac{3^{3/4}}{2} \eta^{1/2}\right)\]
We can also substitute the approximations for $u_0$ and $u_2$ into Eq.~\eqref{eq:intu}. By expanding the result for small $\eta$, we find that the terms of order $\eta^{1/2}$ cancel out. An approximation for the temporal period is given by
\[ T \approx 3- 2\ln(2)  - \frac{3\sqrt{3}}{2}\eta.\]
By combining the approximation for $c$ and that for $T$, we can express $c$ as function of $T$:
\begin{equation}
 c \approx \sqrt{\frac{D_u}{K}} \left(\sqrt{\frac{3}{2}} - \frac{1}{\sqrt{2}} \sqrt{T_0-T}\right). \label{eq:cparabolic}
\end{equation}
Here $T_0 = 3 - 2\ln 2$ is the lowest order approximation of the period of the oscillator itself, computed by taking the integrals over the complete branches of the cubic. This approximation works quite well for waves whose period is close to the period of the oscillating medium (Fig.~\ref{fig:singperturb}e). Such an approximation is useful, since pacemakers often oscillate only slightly faster than the surrounding medium. 

To obtain a numerical value that can be compared to simulations, we need to rescale back from time and space variables $s$ and $y$ to $t$ and $x$. For the speed, this gives
\begin{equation}
c = \frac{K}{\varepsilon^{1/2}} c(v_0). 
\end{equation}
Recall that $K$ is the factor with which we rescale the temporal dynamics in order to obtain a period of exactly $P_o$ for the oscillating medium. This value is equal to $T/P_o$, where $T$ is the period of the unscaled oscillating system and $P_o$ is our chosen period. For small $\varepsilon$, $T \approx T_0 = 3 - 2\ln 2$. 
Using the parabolic approximation Eq.~\eqref{eq:cparabolic} and $K\approx T_0/P_o$, we find in dimensional units that 
\[ c \approx \sqrt{\frac{T_0 D_u}{2 P_o \varepsilon}}\left(\sqrt{3} - \sqrt{T_0 (P_o-P_i)/P_o}\right). \]
Here, we have used $P_i$ for the period of the waves generated by the pacemaker. This is an approximation, since the effective period of the outgoing waves lies in between the natural period of the pacemaker $P_i$ and the period of the outside medium $P_o$. The approximation to the wave speed is shown in red in Fig.~\ref{fig:singperturb}f, compared to the results of numerical simulation. It becomes accurate for small values of $\varepsilon$, as expected.
Note that in this approximation, the fact that the waves are generated by a pacemaker is never used: the speed is obtained by inserting a temporal period into a dispersion relation, and we have assumed that the period of the waves sent out by the pacemaker is equal to the pacemaker's natural period. We will see later that this is justified for large pacemakers. 

\section{Phase reduction} \label{sec:phasered}
\subsection{Phase diffusion equation}

The second approximation method we use is called the phase reduction method. This technique is often used to study an oscillator subject to a weak perturbation. It was pioneered by \citet{winfree_biological_1967} for discrete populations of oscillators. A nice, general introduction to the method is given by \citet{nakao_phase_2016} and a good description in the context of neuronal oscillations can be found in the book by \citet[Ch. 8]{ermentrout_mathematical_2010}. In our case, the perturbation is due to diffusion. The theory for such spatial systems has been extensively discussed by \citet{kuramoto_chemical_1984}, on whose work we mainly base our approach. We recall the main ideas of the derivation here. Consider a system of the form 
 
\begin{equation}
 X_t = F(X) + \eta D X_{xx}, \label{eq:startphasered}
\end{equation}
where $X$ is a vector of variables and $D$ is a matrix with diffusion coefficients. For our models, we would have $X = (u,v)^T$ and $D =
\begin{pmatrix} D_u & 0 \\ 0 & D_v \end{pmatrix}$. The parameter $\eta$ is added to explicitly show that diffusion should have a small effect, even though we later set it to 1. We assume that the homogeneous system governed by $X_t = F(X)$ admits a stable limit-cycle solution $X_0(t)$ with period $T$. We now define the phase $\phi$ along the limit cycle as the time elapsed on the cycle, from an arbitrary starting point. Moroever, we define a phase function $\Phi$ in the basin of attraction of the limit cycle, such that $\Phi(x) = \Phi(y)$ if $\lim_{t\to\infty} |X(t;x) - X(t;y)| = 0$. Here $X(t; x)$ means the solution of the system $X_t = F(X)$, starting from $x$. Note that this implies that $\frac{d}{dt} \Phi(X(t;x)) = 1$, which by the chain rule implies that $\nabla_X \Phi(X) \cdot \frac{dX}{dt} = \nabla_X \Phi(X) \cdot F(X) = 1$.  

We now switch to the spatial context. Let $X(x,t)$ be the vector of variables at position $x$ and time $t$ and $\phi(x, t)$ be the phase of the system, assuming that everywhere the system is in the basin of attraction of the limit cycle. We have $\phi(x, t) = \Phi(X(x, t))$. Then 
\begin{align}
 \phi_t &= \nabla_X\Phi(X) \cdot X_t \notag \\
 &= \nabla_X\Phi(X(x, t)) \cdot (F(X) + \eta D X_{xx}) \notag \\
 &= 1 + \eta \nabla_X\Phi(X(x, t))D X_{xx}.\label{eq:phitbeforeapprox}
\end{align}

To proceed, we need to express the right hand side as function of $\phi$. We assume that the weak perturbations do not push the system away from the limit cycle, but only alter its phase. We therefore approximate $X(x, t) = X_0(\phi(x, t))$. Define the function 
\begin{equation}
 Z(\phi) = \left.\nabla_X\Phi(X)\right|_{X=X_0(\phi)}. \label{eq:defZ}
\end{equation}
This function is called the phase sensitivity function or phase response function. It measures how the phase of the system on the cycle changes under infinitesimally small perturbations. This function is a vector function and has one component per variable of the system. In our case, $Z(\phi) = (Z_u(\phi), Z_v(\phi))$. We also use the approximation
\[ X_{xx}(x, t) \approx \partial_{xx} X_0(\phi(x, t)) = X_0'(\phi)\phi_{xx} + X_0''(\phi) \phi_x^2 . \]

Substituting this and Eq.~\eqref{eq:defZ} into Eq.~\eqref{eq:phitbeforeapprox} gives
\begin{equation}
 \phi_t = 1 + \eta Z(\phi) \cdot D \cdot   (X_0'(\phi)\phi_{xx} + X_0''(\phi) \phi_x^2). 
\end{equation}
The second term is assumed to give a contribution on slow timescales. We average its contribution over one period of the oscillation $T$. This, and setting $\eta$ back to 1 finally leads to
\begin{equation}
 \phi_t = 1 + \alpha \phi_{xx} + \beta \phi_{x}^2 \label{eq:phasediffusion}
\end{equation}
with
\begin{equation}
\begin{aligned}
 \alpha &= \frac{1}{T} \int_0^T Z(\phi) \cdot D \cdot X_0'(\phi) \ud\phi \\
 \beta &= \frac{1}{T} \int_0^T Z(\phi) \cdot D \cdot X_0''(\phi) \ud\phi.
 \end{aligned}
\end{equation}

Eq.~\eqref{eq:phasediffusion} is called the nonlinear phase diffusion equation. An equivalent equation was also derived in a different way by \citet{neu_chemical_1979} and \citet{hagan_target_1981}. As mentioned in the introduction, this equation is often used as the starting point for the analysis of wave patterns, leaving $\alpha$ and $\beta$ unspecified. Since we are interested in obtaining an estimate for the speed of waves in particular systems, we will use computed values of $\alpha$ and $\beta$. 
These values can usually not be obtained analytically. They depend on the limit cycle of the system and on the phase sensitivity function $Z$. To obtain numerical values of $\alpha$ and $\beta$, we numerically compute the limit cycle by simulating the system. Once we have the limit cycle solution, we use the adjoint method to compute the function $Z(\phi)$ \cite[Ch. 8]{ermentrout_mathematical_2010}. For this, we use an adapted version of the algorithm given by \citet{nakao_phase_2016}. This method is established for ODEs, but has only recently been described for delay equations \cite{kotani_adjoint_2012, novicenko_phase_2012}. To obtain $Z$ in the delayed case, we mainly followed \cite{novicenko_phase_2012}.

The equations for $\alpha$ and $\beta$ are vector equations. For example, the one for $\alpha$ reads, written out, 
\[ \alpha = \frac{1}{T} \int_0^T \begin{pmatrix} Z_u(\phi)  &  Z_v(\phi) \end{pmatrix} \cdot \begin{pmatrix} D_u & 0 \\ 0 & D_v \end{pmatrix} \cdot \begin{pmatrix} u_0(\phi) \\ v_0(\phi) \end{pmatrix} \ud\phi. \]

In practice, we compute $\alpha$ as $\alpha_u D_u + \alpha_v D_v$ with
\[ \alpha_u = \frac{1}{T} \int_0^T Z_u(\phi) u_0'(\phi) \ud \phi,\quad\alpha_v= \frac{1}{T} \int_0^T Z_v(\phi) v_0'(\phi) \ud \phi \]
and analogously for $\beta$. This way of writing $\alpha$ and $\beta$ also makes it possible to distinguish between the influence of the diffusion in each of the variables on the wave pattern. 

\begin{figure}
\centering
\includegraphics{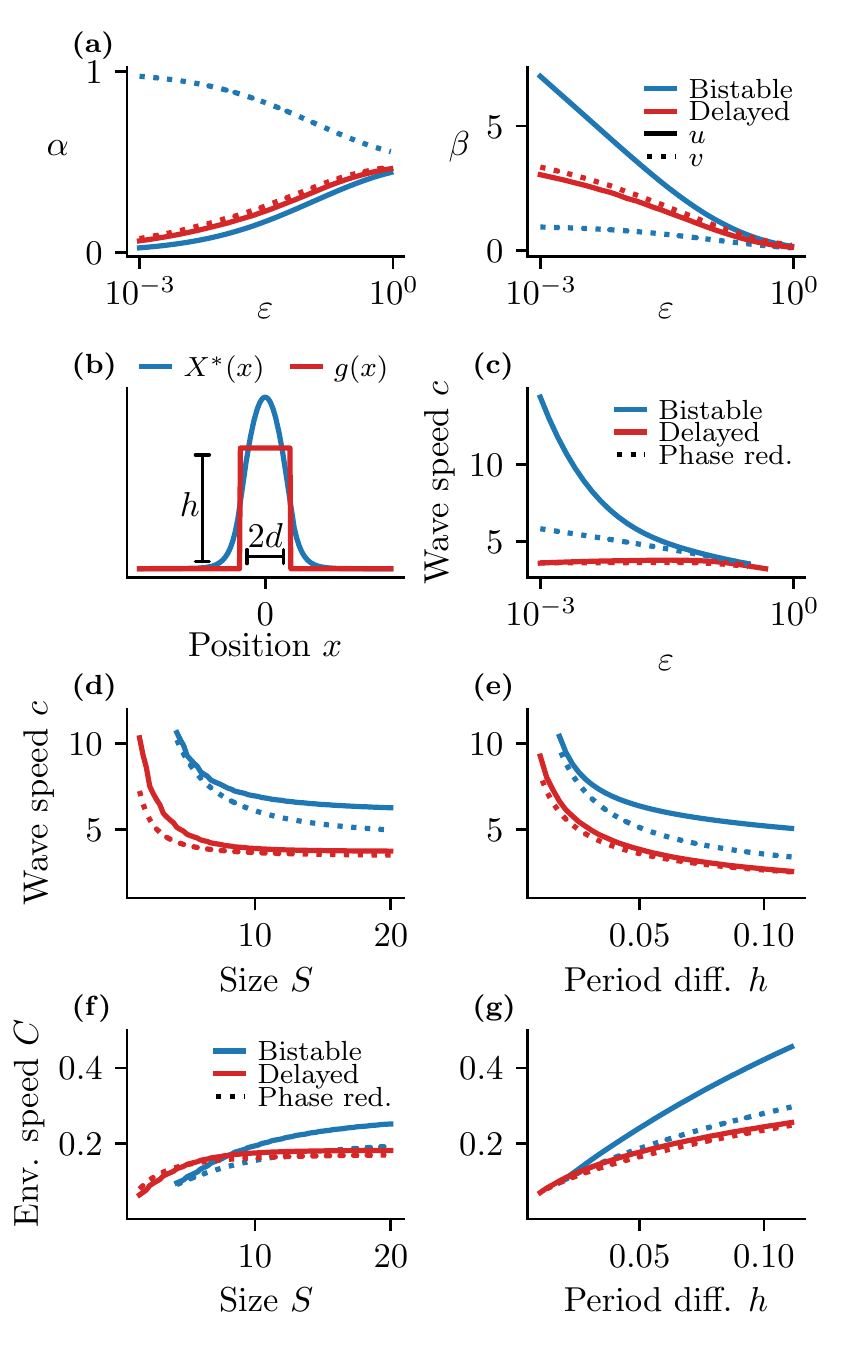}
 \caption{(a) Left: the parameters $\alpha_u$ and $\alpha_v$ as function of $\varepsilon$ for the bistable and the delayed oscillator. Right: values of $\beta_u$ and $\beta_v$. (b) Sketch of the function $g$ which defines the inhomogeneity, and of the spatial eigenfunction $X^*$ associated to the largest eigenvalue $\omega^*$. (c) Wave speed as function of $\varepsilon$ for the bistable and the delayed oscillator: numerical simulation (solid line) and approximation through phase reduction method (dotted line). (d,e) Wave speed as function of pacemaker size (d) and relative period difference (e) for bistable and delayed oscillator: numerical simulation (solid line) compared to phase reduction approximation (dotted line). (f,g) Envelope speed as function of size and period difference. In panels d-g, $\varepsilon=0.01$. In panels d and f, $P_i=9.5$. In panels e and g, $S=20$. }
 \label{fig:phasered1}
\end{figure}

In Fig.~\ref{fig:phasered1}a we show how $\alpha$ and $\beta$ change with the timescale separation parameter for our two models, separately for the $u$ and $v$ variable. For large $\varepsilon$ all values converge: all $\alpha$'s seem to converge to $1/2$ and $\beta$ goes to zero. One explanation for this lies in the fact that for larger values of $\varepsilon$, both the bistable and delayed oscillator are small perturbations of a harmonic oscillator. This can be seen in the phase plane and time series (Fig.~\ref{fig:setup}), and analytically as follows. By setting $s = \varepsilon^{-1/2} t$ and $\tilde v =\varepsilon^{-1/2} v$, the (space-independent) equations for the bistable oscillator become
\begin{equation}
 \begin{aligned}
  u_s &= \tilde v - \varepsilon^{-1/2} f(u) \\
  \tilde v_s &= -u
 \end{aligned}
\end{equation}
and for the delayed oscillator they are 
\begin{equation}
 \begin{aligned}
  u_s &= \tilde v\left(s-\varepsilon^{-1/2} \tau\right) - \varepsilon^{-1/2} \tilde f(u) \\
  \tilde v_s &= -u.
 \end{aligned}
\end{equation}
For large $\varepsilon$, or equivalently small $\varepsilon^{-1/2}$, both systems reduce to $u_s = \tilde v, \tilde v_s = -u$ to leading order, which has sinusoidal solutions. This may explain why for larger $\varepsilon$, both oscillators show similar values of $\alpha$ and $\beta$. However, for computing the wave speed this observation is less useful, since for larger $\varepsilon$ clear wave propagation is often not seen \cite{rombouts_synchronizing_2020}. 

The plots in Fig.~\ref{fig:phasered1}a also show that for the delayed model there is only very little difference between $\alpha_u$ and $\alpha_v$, and similarly for $\beta$. For the bistable oscillator there are clear differences: $\alpha_u$ increases with $\varepsilon$ whereas $\alpha_v$ decreases. The value of $\beta_u$ is much larger than that of $\beta_v$ for the bistable oscillator. This hints at the relative importance of the diffusion in $u$ and $v$: for the delayed oscillator their contribution is similar, whereas for the bistable oscillator they are different. As we will see below, the value of $\beta$ is the main determinant of the wave speed for larger pacemakers. Since $\beta_u > \beta_v$ for the bistable system, diffusion in the $u$-variable is more important. This was also the case in the singular perturbation result, where only $D_u$ appears in the expression for the wave speed. This corresponds to our earlier numerical results on the influence of diffusion \cite{rombouts_synchronizing_2020}. 

We have computed values for $\alpha$ and $\beta$ for the non-time-scaled oscillator. To compare with the simulations, in which we rescale time such that the medium always has period $P_o$, we need to rescale $\beta$. A calculaton shows that there, we need to use $\beta \times \frac{P}{P_o}$, where $P$ is the period of the unscaled oscillator. 

\subsection{Solution of the phase diffusion equation in the presence of an inhomogeneity}
The nonlinear phase diffusion equation can be derived in the presence of an inhomogeneity, such as a pacemaker, which is also considered a small perturbation \cite[Ch. 6]{kuramoto_chemical_1984}.  In our case, this modification leads to the equation
\begin{equation}
 \phi_t = 1 + \alpha \phi_{xx} + \beta \phi_x^2 + g(x), \label{eq:phasediffusionwithg}
\end{equation}
with 
\begin{equation}
 g(x) = \begin{cases}
         \frac{P_o - P_i}{P_i} & \text{if}\ |x-L/2| < S/2 \\
         0 & \text{elsewhere}.
        \end{cases}
\end{equation}
This equation describes that in the pacemaker region, the speed by which the phase changes is $1+(P_o-P_i)/P_i$ whereas in the outside medium it is $1$.

This equation allows solutions with waves traveling outward from the pacemaker. To see this, we first set $\psi = \phi - t$ and then perform the Cole-Hopf transform $Y = e^{\frac{\beta}{\alpha} \psi}$. The equation for $Y$ is
\begin{equation}
 Y_t = \alpha Y_{xx} + \frac{\beta}{\alpha}g(x) Y,
\end{equation}
which is linear. Using separation of variables: $Y(x, t) = X(x)T(t)$, we find
\begin{align}
 T' &= \omega T\\
 \alpha X'' + \frac{\beta}{\alpha} g(x) X &= \omega X. \label{eq:X}
\end{align}
Here, $\omega$ is a constant. The solution for $Y$ is then of the form
\[ Y(x, t) = \sum_n X_n(x) e^{\omega_n t}. \]
The $\omega_n$ are the eigenvalues of the equation for $X$. At large times, the behavior will be dominated by the term with the largest eigenvalue. If we call this eigenvalue $\omega^*$, and its associated spatial eigenfunction $X^*$, we find for $\phi$ at large times:
 \begin{align}
  \phi(x, t) &= t + \psi(x, t) \notag \\
  &= t + \frac{\alpha}{\beta} \ln Y(x,t) \notag \\
  &\approx t + \frac{\alpha}{\beta} \ln \left( X^*(x) e^{\omega^* t} \right) \notag \\
  &=(1+\frac{\alpha}{\beta}\omega^*)t + \frac{\alpha}{\beta}\ln X^*(x). \label{eq:philarget}
 \end{align}

We now show how to find $\omega^*$ and $X^*$ in the case of a piecewise-constant inhomogeneity $g$. For simplicity, we assume that the domain is infinite and that the pacemaker is situated at $x=0$. This is justified since we are interested in situations where the pacemaker is much smaller than the domain. We also define $d = S/2$ and $h = (P_o - P_i)/P_i$ (Fig.~\ref{fig:phasered1}b). 

The equation for $X$, Eq.~\eqref{eq:X}, reads written out
\begin{equation}
\begin{aligned}
 \alpha X'' &= (\omega - \frac{\beta}{\alpha} h) X & &\text{for}\ |x| < d \\
 \alpha X'' &= \omega X & &\text{elsewhere}. 
\end{aligned} 
\end{equation}
Interestingly, this equation is equivalent to the Schr\"odinger equation for a particle in a finite square potential well, which is treated in many quantum mechanics textbooks. 
The equation can be solved exactly on the different regions in terms of exponentials and trigonometric functions. By imposing continuity of $X$ and $X'$ and the conditions that $X(x) > 0$ and that $X(x)\to0$ as $x\to\pm\infty$, the solution we retain is a multiple of 
\begin{equation}
\begin{aligned}
X(x) &= \cos\left(\sqrt{\frac{\frac{\beta}{\alpha}h-\omega}{\alpha}}x\right) & &\text{for}\ |x| < d \\
 X(x) &= e^{-\sqrt{\frac{\omega}{\alpha}} |x|} & &\text{elsewhere}. 
\end{aligned} 
\end{equation}
Such a solution is shown in Fig.~\ref{fig:phasered1}b. The values of $\omega$ for which this solution is possible satisfy $0< \omega < \frac{\beta}{\alpha} h$ and
\begin{equation}
 \tan\left(\sqrt{\frac{\frac{\beta}{\alpha}h - \omega}{\alpha}} d\right) = \sqrt{\frac{\omega}{\beta h/\alpha - \omega}}. \label{eq:omega}
\end{equation}

For large $|x|$, the function $X$ decays exponentially as $e^{-\sqrt{\frac{\omega}{\alpha}} |x|}$. We can substitute the expression for $X(x)$ at large distances from the origin into Eq.~ \eqref{eq:philarget} to obtain, for large $t$ and $x$:
\begin{equation}
 \phi(x, t) \approx (1+\frac{\alpha}{\beta}\omega^*)t - \frac{\alpha}{\beta}\sqrt{\frac{\omega^*}{\alpha}}|x| \label{eq:phiwavelarget}
\end{equation}
This solution corresponds to a traveling wave with speed
\begin{equation}
 c = \frac{1 + \frac{\alpha}{\beta} \omega^*}{\frac{\alpha}{\beta} \sqrt{\frac{\omega^*}{\alpha}}}. \label{eq:c}
\end{equation}
It is also clear that this solution oscillates a factor $1 + \frac{\alpha}{\beta}\omega^*$ faster than the homogeneous oscillation. The effective period of the wave pattern is thus
\begin{equation}
 P_t = \frac{P_o}{1 + \frac{\alpha}{\beta}\omega^*}. \label{eq:Ptphase}
\end{equation}

The calculation above gives us a procedure to calculate the wave speed, given the parameters of the oscillator and the diffusion coefficients, from which we calculate $\alpha$ and $\beta$. We then combine these values with $h$ and $d$, which depend only on the pacemaker itself, to numerically solve Eq.~\eqref{eq:omega} for the largest $\omega$. This can then be used to compute the wave speed. 

The approximation we obtain in this way is very accurate for the delayed oscillator, and accurate for larger values of $\varepsilon$ for the bistable oscillator (Fig.~\ref{fig:phasered1}c). Unlike the singular perturbation method, the approximation for $c$ obtained through the phase equation takes into account properties of the pacemaker such as size and frequency. Fig.~\ref{fig:phasered1}d,e shows how the wave speed depends on these features. Large pacemakers lead to slower waves, and the speed converges to a constant for very large pacemakers. A larger period difference between pacemaker and surroundings also leads to a slower wave. The curves do not continue to $S=0$ and $h=0$, since for very small pacemakers or pacemakers with a small period difference, we do not observe linearly propagating waves. 

In our previous paper, we showed that the envelope speed $C$ is related to the wave speed $c$ through the formula \cite{rombouts_synchronizing_2020}
\begin{equation}
 C = \frac{P_o-P_t}{P_o} c. \label{eq:C}
\end{equation}
Generally, factors that increase the wave speed, such as a lower $\varepsilon$ in the bistable oscillator, have a similar effect on the envelope speed. However, size and period difference of the pacemaker regions have a more intricate effect (Fig.~\ref{fig:phasered1}f,g): a larger pacemaker sends out waves with a lower speed, but the envelope speed is higher. Similarly, pacemakers with a higher period difference with the environment send out slower waves, but have a higher envelope speed. This difference in the behavior of wave and envelope speed can be attributed to changes in $P_t$, the effective period of the waves. This period lies between $P_i$, the natural period of the pacemaker, and $P_o$, the period of the outside medium. This effective period is influenced by diffusion strength and the size of the pacemaker. For example, larger pacemakers are less influenced by the outside medium, and the effective period $P_t$ decreases towards $P_i$ as the pacemaker's size increases. This means that $P_t - P_o$ becomes larger with size, whereas $c$ decreases with size (Fig.~\ref{fig:phasered1}d).  The effect on $P_t$ dominates the decrease in $c$, which results in a net increase in envelope speed computed through Eq.~\eqref{eq:C}.


We can better understand the dependencies of the wave speed on features of the pacemaker by further considering Eq.~\eqref{eq:omega}. First, some insight into the behavior of the roots can be gained by transforming the equation into an easier one. This approach is identitical to the one explained by \citet{de_alcantara_bonfim_exact_2005} in the context of quantum mechanics. By setting $\tilde \omega = \frac{\alpha}{\beta h} \omega$, the equation becomes
\[ \tan\left(\sqrt{\frac{1-\tilde\omega}{\rho}}\right) = \sqrt{\frac{\tilde\omega}{1-\tilde\omega}},\]
where $\rho = \alpha^2/(\beta h d^2)$. Now set $\tilde\omega = 1- \rho x^2$, such that we have
\[ \tan|x| = \sqrt{\frac{1}{\rho x^2} - 1}.\]
Squaring, using a trigonometric identity and taking the square root again finally leads to the simpler equation
\begin{equation}
 \cos(x) = \sqrt{\rho} x. \label{eq:rhox}
\end{equation}
Recall that, to describe the outgoing waves, we are interested in the largest value of $\omega$ that is smaller than $\beta h/\alpha$. This means we look for the smallest value of $x < 1/\sqrt{\rho}$ that is a solution of Eq.~\eqref{eq:rhox}. This value can be found as the intersection of the cosine function on the interval $[0, \pi/2]$ with the straight line $\sqrt{\rho} x$ (Fig.~\ref{fig:phasered_approx}a). The equation cannot be solved exactly, but a very good approximation can be obtained by replacing the cosine by a parabola and using the quadratic formula \cite{de_alcantara_bonfim_exact_2005}. This formula is quite unwieldy and does not provide much additional insight. 
The graphical approach does show, however, that there always is at least one positive $\omega$ that satisfies the equation. This is not the case if the pacemaker problem is solved in three dimensions, as done by \citet{kuramoto_chemical_1984}: in that case, a positive eigenvalue does not necessarily exist, and there is a threshold condition on the size and frequency of the pacemaker for outgoing wave patterns to occur.

\subsection{Approximations for small and large pacemakers}

\begin{figure}
 \centering
 \includegraphics{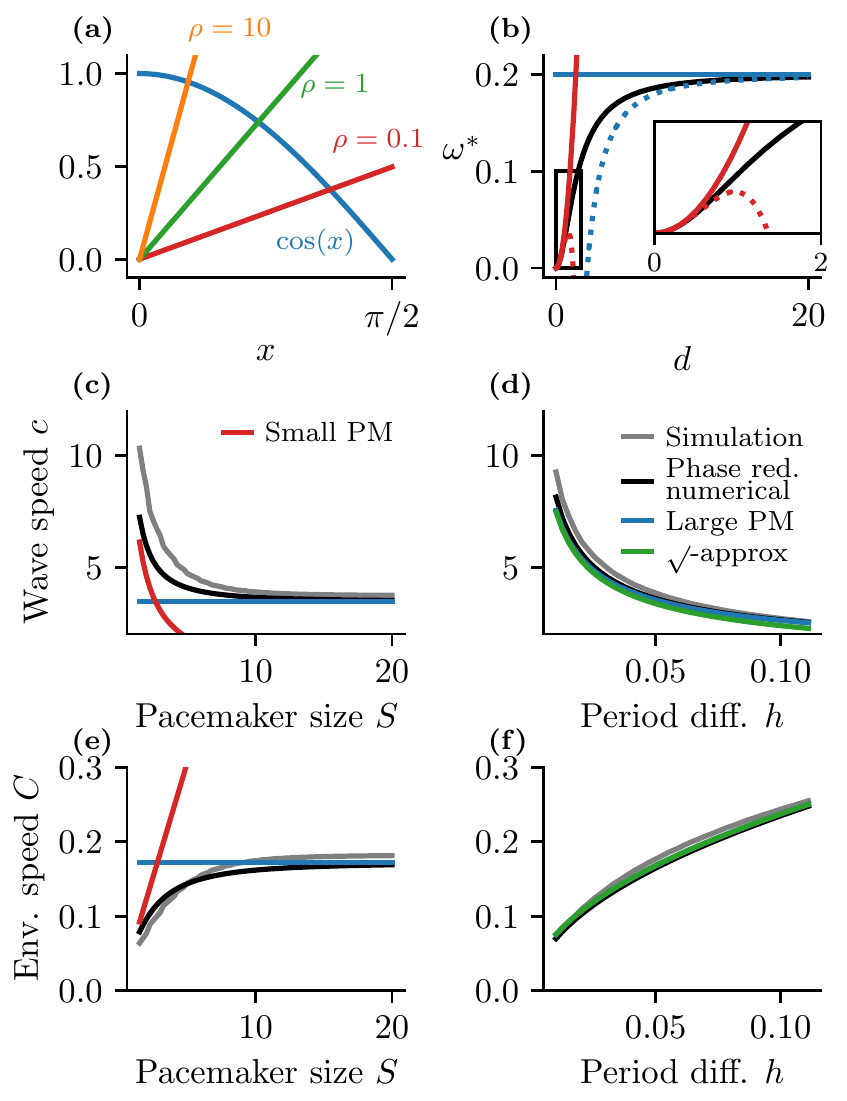}
 \caption{(a) Graphical representation of Eq.~\eqref{eq:rhox}. The intersection of the curves given by $\cos x$ and $\sqrt{\rho} x$ determines the largest eigenvalue $\omega^*$. (b) Asymptotic approximations of $\omega^*$ as root of Eq.~\eqref{eq:omega} for small and large pacemakers. The solid black line is the numerically obtained solution. The blue lines are approximations for large $d$, the red lines are the small-$d$ approximation. The solid and dotted colored lines are the approximations obtained by keeping one and two terms in the expansion for $\omega^*$ respectively. The inset axis shows a zoom for small $d$. Computed with $\alpha=1/2, \beta=2, h=0.05$. (c) and (d): Approximations to the wave speed as function of pacemaker size and period difference, for the delayed oscillator with $\varepsilon=0.01$. On the left, $P_i=9.5$, on the right $S=20$. Because the size is large, only the large pacemaker approximation is shown in panel d. (e,f) Envelope speed as function of size and relative period difference.}
 \label{fig:phasered_approx}
\end{figure}

The solution of Eq.~\eqref{eq:omega} can be approximated in certain limits. We consider here the limits of small and large pacemakers. In the graphical representation in Fig.~\ref{fig:phasered_approx}a, the value of $\rho$ determines the solution. Since $\rho=\alpha^2/(\beta h d^2)$, we expect that the value of $d$ is an important parameter to vary. Using perturbation theory --- and with some help from computer algebra --- we find two asymptotic expressions for $\omega^*$: 
\begin{align}
 \omega^* &= \frac{\beta^2 h^2}{\alpha^3} d^2 - \frac{4}{3} \frac{\beta^3 h^3}{\alpha^5} d^4 + \mathcal{O}(d^6) & &\text{for}\ d\to0 \\
 \omega^* &= \frac{\beta h}{\alpha} - \frac{\alpha \pi^2}{4} \frac{1}{d^2} + \mathcal{O}(1/d^4)& &\text{for}\ d\to\infty
\end{align}

These approximations are shown in Fig~\ref{fig:phasered_approx}b. The approximation for large $d$ (blue curves) works well over a reasonably large interval, but the approximation for small $d$ (red curves) is only accurate for very small pacemakers. In this regime, however, the waves do not propagate linearly outward anymore from the pacemaker \cite{rombouts_synchronizing_2020}.  Perhaps this is related to the fact that for small $d$, $\omega^*$ is also very small, which means that the contribution of the term with $e^{\omega^* t}$ does not dominate until very long times, whereas in our simulations we always look at the transient behavior. 

By substituting the asymptotic expressions in the formula for the wave speed (Eq.~\eqref{eq:c}), we find to leading order in $d$:
\begin{align}
 c &\approx \frac{\alpha h}{d} & &\text{for}\ d\to0\\
 c &\approx \sqrt{\frac{\beta}{h}} + \sqrt{\beta h}& &\text{for}\ d\to\infty
\end{align} 
This shows that the speed decreases with pacemaker size, and goes to a constant value of large pacemakers (Fig.~\ref{fig:phasered_approx}c). As expected, the approximation for small pacemakers is only good for very small $d$, and quickly deviates from the numerically obtained solution. For small and large pacemakers, the dependence on the period difference $h$ is different, and it is notable that for small pacemakers the parameter $\alpha$ determines the speed, but for large pacemakers it is $\beta$. 

The period difference $h$ between pacemaker and outside medium is typically small. If we consider this and keep only the first term for the large-pacemaker approximation, we find a simple approximative formula for the wave speed:
\begin{equation}
  c \approx \sqrt{\beta / h}. \label{eq:csqrt}
\end{equation}
This is already a very good approximation, as can be seen in Fig.~\ref{fig:phasered_approx}d. For large pacemakers, Eq.~\eqref{eq:Ptphase} leads to $P_t \approx P_o / (1+h) = P_i$: if the pacemaker is large, the period of the pattern of outgoing waves is nearly the same as the period of the pacemaker region, if no spatial effects would be present. Intuitively this makes sense: we expect that larger pacemakers feel less effect from their environment. 

The formula $c \approx \sqrt{\beta / h}$, valid for large pacemakers and small period differences, also provides an analytical argument for the behavior of the envelope speed as function of $h$. The value of $h$ is $\frac{P_o-P_i}{P_i}$, which for small differences in period is nearly the same as $\frac{P_o-P_i}{P_o}$. By combining Eq.~\eqref{eq:C} with Eq.~\eqref{eq:csqrt}, we find that the envelope speed $C$ is approximately $hc = \sqrt{\beta h}$. This shows that, whereas $c$ decreases with $h$, the envelope speed increases. 

\section{Discussion} \label{sec:discussion}

\begin{figure}
 \includegraphics{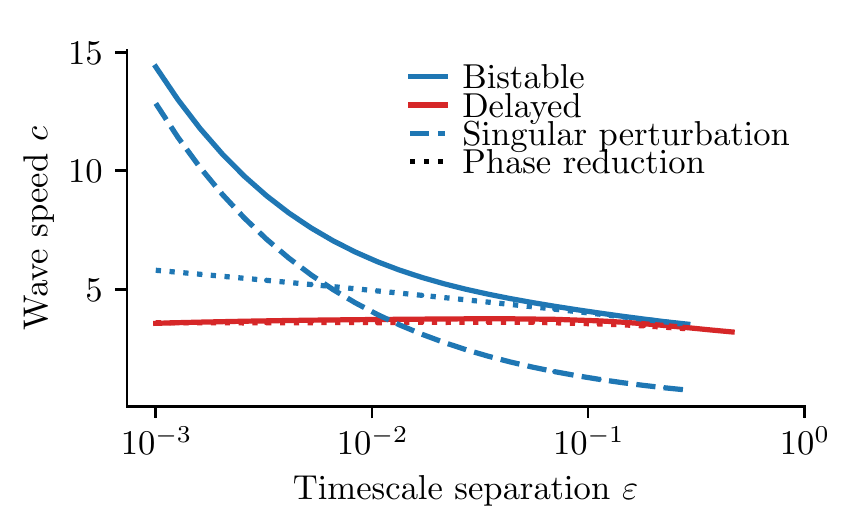}
 \caption{Wave speed as function of $\varepsilon$ with different approximations. }
 \label{fig:alltogether}
\end{figure}

In this paper, we have used two well-known mathematical methods to describe traveling waves and obtained estimates of the wave speed of pacemaker-generated waves. Whereas previous work used these methods to explain qualitative aspects of wave patterns, our goal was to see whether these methods can give a quantitative estimate of the wave speed, and compare these to explicit numerical simulations of the PDEs. 
We did this for two different oscillatory systems, one based on an underlying bistability and one based on a time-delayed negative feedback loop. These systems show very similar time series, but their spatially extended version shows markedly different behavior. This is especially clear when considering the influence of the timescale separation parameter $\varepsilon$ (Fig.~\ref{fig:alltogether}). This parameter has an important effect on the speed of traveling waves for the bistable oscillator, but not for the delayed oscillator. The behavior at low $\varepsilon$ is, for the bistable oscillator, well captured by the singular perturbation method. In this method, traveling waves are considered as propagating bistable fronts, whose speed determines the speed of the waves. Such fronts are not present in the delayed system, which may explain the difference. 

From the singular perturbation calculation, we see that the speed scales as $\sqrt{D_u / \varepsilon}$. This is reminiscent of Luther's formula, which states that the speed of a wave can be calculated as $c_0\sqrt{D k}$, where $D$ is the diffusion constant, $k$ an autocatalysis rate and $c_0$ a dimensionless constant of order unity. This formula appears in Luther's original German article \cite{luther_raumliche_1906} (translated by \citet{arnold_propagation_1987} and discussed by \citet{showalter_luthers_1987}). Luther's formula can be derived for a propagating front where a stable state invades an unstable state. It has also been derived for the Fisher-KPP equation, which couples logistic growth with diffusion and is often used in population dynamics. 
The singular perturbation shows that the use of Luther's formula is also justified for waves with an underlying bistability if $1/\varepsilon$ is treated as the autocalysis rate. This can be done since jumps between the two branches of the cubic nullcline happen on a timescale of $\varepsilon$. Nevertheless, our analysis shows that the formula is not applicable if the waves originate from a time-delayed system coupled by diffusion. The formula should thus be used with care. 

The faster waves for the bistable oscillator suggest that many real biological systems where waves play a functional role are based on an underlying bistability. Systems with such a bistability can also be excitable, rather than oscillatory. Excitable media can equally sustain pacemaker-generated waves, which can be analyzed using the singular perturbation method. We expect therefore that our results for the bistable oscillator may also be relevant for excitable media. 

The dependency of the wave speed on the timescale separation $\varepsilon$, shown in Fig.~\ref{fig:alltogether}, suggests that there are two regimes for the bistable oscillator. For low $\varepsilon$, the singular perturbation method applies and waves can be treated as traveling fronts. For higher $\varepsilon$, the dependence of the speed on $\varepsilon$ is weaker. In this regime, the oscillations are no longer of relaxation type but become closer to sinusoidal oscillations. Here, the phase approximation method gives an accurate estimate of the wave speed. This method breaks down for small $\varepsilon$, likely because of the sharp jumps in this regime where the spatial derivatives $u_{xx}$ and $v_{xx}$ are large. This does not correspond with the assumption of the phase reduction method that diffusion is a small perturbation. It would be interesting to find an analytical method to describe waves in the transition region with intermediate values of $\varepsilon$. 

It is curious that the phase reduction method gives such a good estimate for the wave speed over the whole range of $\varepsilon$ for the delayed oscillator, since for smaller values of $\varepsilon$ the waves in this system also show sharp jumps. To our knowledge, the phase reduction method has not been applied to delayed systems in a spatial context. From a theoretical point of view, it would be very interesting to investigate whether in other oscillatory systems based on time delays, the phase reduction method is also accurate. The derivation of the phase equation is much more recent for delay equations than for ODEs \cite{novicenko_phase_2012, kotani_adjoint_2012}, and is still studied \cite{kotani_nonlinear_2020}. 

In our analysis of the phase diffusion equation with a pacemaker, we have discussed the solution in the one-dimensional case, analogously to the computation performed by \citet{kuramoto_chemical_1984} for three dimensions. Interestingly, this computation was already performed by others in the context of quantum mechanics \cite{de_alcantara_bonfim_exact_2005}. In order to obtain an estimate for the wave speed, we had to solve an equation for the largest eigenvalue $\omega^*$. By using asymptotic expansions, we derived simple formulas for the wave speed in the limit of small and large pacemakers, of which the large pacemaker one is the most relevant. A possible extension of this part of the study could look at higher dimensions. In 2D for example, not only the size of the pacemaker but also its shape plays a role. In 2D, square, disk and ring-shaped pacemakers have already been studied \cite{mahara_ring-shaped_2000,stich_target_2006}.

Often, multiple pacemakers are present in a single medium. In such a situation, each pacemaker may send out waves, until after a long time one pacemaker has entrained the others. This has been shown theoretically already by \citet{hagan_target_1981} and \citet{kuramoto_chemical_1984}. The phase approximation can be used to study such systems: if the distance between different pacemakers is large enough, the solutions to Eq.~\eqref{eq:X} are localized around each pacemaker and can be computed as we did here. In those cases, after a long time the pacemaker with largest associated $\omega^*$ will entrain the whole medium \cite{hagan_target_1981, kuramoto_chemical_1984}.  Since $\omega^*$ depends both on $h$ and on $d$, the size and frequency difference of a pacemaker both contribute to whether it will dominate. For equally sized pacemakers, the one with the largest period difference with the surroundings will win out in the end, but a smaller period difference can be compensated by a larger size. We have numerically investigated this in a previous paper \cite{nolet_synchronization_2020}. It would be interesting to use the approach we took here, using the phase equation, to make quantitative predictions on how exactly two pacemakers interact, and compare those to numerical experiments. 

The heterogeneity $g(x)$ in the phase equation (Eq.~\eqref{eq:phasediffusionwithg}) does not have to be localized or piecewise constant, even though this was convenient since it allowed an analytical solution for the eigenfunctions $X(x)$. \citet{tonjes_perturbation_2009} investigated the synchronization properties of the phase diffusion equation with a random frequency distribution using perturbation theory. Again, it would be interesting to perform a similar study for a concrete reaction-diffusion system, and compare the results from a phase description to numerical solutions of the original set of PDEs. 

The mathematical methods to describe wave patterns were first developed decades ago, usually inspired by experimental observations in chemical reactions. In those systems, a pacemaker arises through a scratch or dust particle in the reaction dish, almost as a side effect. Nowadays, more and more studies highlight the importance of oscillations and waves in biological systems, where these waves can provide essential functions to an organism. We expect that mathematical modeling of such waves still has an important role to play in describing and understanding these spatial structures.

\section*{Acknowledgments}
 
We thank Felix Nolet and Arno Vanderbeke for comments on the manuscript. This work was supported by the Research Foundation Flanders (FWO) with individual support to J.R. and project support to L.G. (Grant No. GOA5317N) and the KU Leuven
Research Fund (Grant No. C14/18/084).

\bibliographystyle{apsrev4-1}

\begin{thebibliography}{41}%
\makeatletter
\providecommand \@ifxundefined [1]{%
 \@ifx{#1\undefined}
}%
\providecommand \@ifnum [1]{%
 \ifnum #1\expandafter \@firstoftwo
 \else \expandafter \@secondoftwo
 \fi
}%
\providecommand \@ifx [1]{%
 \ifx #1\expandafter \@firstoftwo
 \else \expandafter \@secondoftwo
 \fi
}%
\providecommand \natexlab [1]{#1}%
\providecommand \enquote  [1]{``#1''}%
\providecommand \bibnamefont  [1]{#1}%
\providecommand \bibfnamefont [1]{#1}%
\providecommand \citenamefont [1]{#1}%
\providecommand \href@noop [0]{\@secondoftwo}%
\providecommand \href [0]{\begingroup \@sanitize@url \@href}%
\providecommand \@href[1]{\@@startlink{#1}\@@href}%
\providecommand \@@href[1]{\endgroup#1\@@endlink}%
\providecommand \@sanitize@url [0]{\catcode `\\12\catcode `\$12\catcode
  `\&12\catcode `\#12\catcode `\^12\catcode `\_12\catcode `\%12\relax}%
\providecommand \@@startlink[1]{}%
\providecommand \@@endlink[0]{}%
\providecommand \url  [0]{\begingroup\@sanitize@url \@url }%
\providecommand \@url [1]{\endgroup\@href {#1}{\urlprefix }}%
\providecommand \urlprefix  [0]{URL }%
\providecommand \Eprint [0]{\href }%
\providecommand \doibase [0]{http://dx.doi.org/}%
\providecommand \selectlanguage [0]{\@gobble}%
\providecommand \bibinfo  [0]{\@secondoftwo}%
\providecommand \bibfield  [0]{\@secondoftwo}%
\providecommand \translation [1]{[#1]}%
\providecommand \BibitemOpen [0]{}%
\providecommand \bibitemStop [0]{}%
\providecommand \bibitemNoStop [0]{.\EOS\space}%
\providecommand \EOS [0]{\spacefactor3000\relax}%
\providecommand \BibitemShut  [1]{\csname bibitem#1\endcsname}%
\let\auto@bib@innerbib\@empty
\bibitem [{\citenamefont {Rombouts}\ and\ \citenamefont
  {Gelens}(2020)}]{rombouts_synchronizing_2020}%
  \BibitemOpen
  \bibfield  {author} {\bibinfo {author} {\bibfnamefont {J.}~\bibnamefont
  {Rombouts}}\ and\ \bibinfo {author} {\bibfnamefont {L.}~\bibnamefont
  {Gelens}},\ }\href {\doibase 10.1103/PhysRevResearch.2.043038} {\bibfield
  {journal} {\bibinfo  {journal} {Physical Review Research}\ }\textbf {\bibinfo
  {volume} {2}},\ \bibinfo {pages} {043038} (\bibinfo {year}
  {2020})}\BibitemShut {NoStop}%
\bibitem [{\citenamefont {Nov{\'a}k}\ and\ \citenamefont
  {Tyson}(2008)}]{novak_design_2008}%
  \BibitemOpen
  \bibfield  {author} {\bibinfo {author} {\bibfnamefont {B.}~\bibnamefont
  {Nov{\'a}k}}\ and\ \bibinfo {author} {\bibfnamefont {J.~J.}\ \bibnamefont
  {Tyson}},\ }\href {\doibase 10.1038/nrm2530} {\bibfield  {journal} {\bibinfo
  {journal} {Nature Reviews Molecular Cell Biology}\ }\textbf {\bibinfo
  {volume} {9}},\ \bibinfo {pages} {981} (\bibinfo {year} {2008})}\BibitemShut
  {NoStop}%
\bibitem [{\citenamefont {Taylor}(2002)}]{taylor_mechanism_2002}%
  \BibitemOpen
  \bibfield  {author} {\bibinfo {author} {\bibfnamefont {A.~F.}\ \bibnamefont
  {Taylor}},\ }\href {\doibase 10.3184/007967402103165414} {\bibfield
  {journal} {\bibinfo  {journal} {Progress in Reaction Kinetics and Mechanism}\
  }\textbf {\bibinfo {volume} {27}},\ \bibinfo {pages} {247} (\bibinfo {year}
  {2002})}\BibitemShut {NoStop}%
\bibitem [{\citenamefont {Jakubith}\ \emph {et~al.}(1990)\citenamefont
  {Jakubith}, \citenamefont {Rotermund}, \citenamefont {Engel}, \citenamefont
  {{von Oertzen}},\ and\ \citenamefont {Ertl}}]{jakubith_spatiotemporal_1990}%
  \BibitemOpen
  \bibfield  {author} {\bibinfo {author} {\bibfnamefont {S.}~\bibnamefont
  {Jakubith}}, \bibinfo {author} {\bibfnamefont {H.~H.}\ \bibnamefont
  {Rotermund}}, \bibinfo {author} {\bibfnamefont {W.}~\bibnamefont {Engel}},
  \bibinfo {author} {\bibfnamefont {A.}~\bibnamefont {{von Oertzen}}}, \ and\
  \bibinfo {author} {\bibfnamefont {G.}~\bibnamefont {Ertl}},\ }\href {\doibase
  10.1103/PhysRevLett.65.3013} {\bibfield  {journal} {\bibinfo  {journal}
  {Physical Review Letters}\ }\textbf {\bibinfo {volume} {65}},\ \bibinfo
  {pages} {3013} (\bibinfo {year} {1990})}\BibitemShut {NoStop}%
\bibitem [{\citenamefont {Wolff}\ \emph {et~al.}(2004)\citenamefont {Wolff},
  \citenamefont {Stich}, \citenamefont {Beta},\ and\ \citenamefont
  {Rotermund}}]{wolff_laser-induced_2004}%
  \BibitemOpen
  \bibfield  {author} {\bibinfo {author} {\bibfnamefont {J.}~\bibnamefont
  {Wolff}}, \bibinfo {author} {\bibfnamefont {M.}~\bibnamefont {Stich}},
  \bibinfo {author} {\bibfnamefont {C.}~\bibnamefont {Beta}}, \ and\ \bibinfo
  {author} {\bibfnamefont {H.~H.}\ \bibnamefont {Rotermund}},\ }\href {\doibase
  10.1021/jp0498015} {\bibfield  {journal} {\bibinfo  {journal} {The Journal of
  Physical Chemistry B}\ }\textbf {\bibinfo {volume} {108}},\ \bibinfo {pages}
  {14282} (\bibinfo {year} {2004})}\BibitemShut {NoStop}%
\bibitem [{\citenamefont {Bretschneider}\ \emph {et~al.}(2016)\citenamefont
  {Bretschneider}, \citenamefont {Othmer},\ and\ \citenamefont
  {Weijer}}]{bretschneider_progress_2016}%
  \BibitemOpen
  \bibfield  {author} {\bibinfo {author} {\bibfnamefont {T.}~\bibnamefont
  {Bretschneider}}, \bibinfo {author} {\bibfnamefont {H.~G.}\ \bibnamefont
  {Othmer}}, \ and\ \bibinfo {author} {\bibfnamefont {C.~J.}\ \bibnamefont
  {Weijer}},\ }\href {\doibase 10.1098/rsfs.2016.0047} {\bibfield  {journal}
  {\bibinfo  {journal} {Interface Focus}\ }\textbf {\bibinfo {volume} {6}},\
  \bibinfo {pages} {20160047} (\bibinfo {year} {2016})}\BibitemShut {NoStop}%
\bibitem [{\citenamefont {Chang}\ and\ \citenamefont
  {Ferrell}(2013)}]{chang_mitotic_2013}%
  \BibitemOpen
  \bibfield  {author} {\bibinfo {author} {\bibfnamefont {J.~B.}\ \bibnamefont
  {Chang}}\ and\ \bibinfo {author} {\bibfnamefont {J.~E.}\ \bibnamefont
  {Ferrell}, \bibfnamefont {Jr.}},\ }\href {\doibase 10.1038/nature12321}
  {\bibfield  {journal} {\bibinfo  {journal} {Nature}\ }\textbf {\bibinfo
  {volume} {500}},\ \bibinfo {pages} {603} (\bibinfo {year}
  {2013})}\BibitemShut {NoStop}%
\bibitem [{\citenamefont {Nolet}\ \emph
  {et~al.}(2020{\natexlab{a}})\citenamefont {Nolet}, \citenamefont
  {Vandervelde}, \citenamefont {Vanderbeke}, \citenamefont {Pi{\~n}eros},
  \citenamefont {Chang},\ and\ \citenamefont {Gelens}}]{nolet_nuclei_2020}%
  \BibitemOpen
  \bibfield  {author} {\bibinfo {author} {\bibfnamefont {F.~E.}\ \bibnamefont
  {Nolet}}, \bibinfo {author} {\bibfnamefont {A.}~\bibnamefont {Vandervelde}},
  \bibinfo {author} {\bibfnamefont {A.}~\bibnamefont {Vanderbeke}}, \bibinfo
  {author} {\bibfnamefont {L.}~\bibnamefont {Pi{\~n}eros}}, \bibinfo {author}
  {\bibfnamefont {J.~B.}\ \bibnamefont {Chang}}, \ and\ \bibinfo {author}
  {\bibfnamefont {L.}~\bibnamefont {Gelens}},\ }\href {\doibase
  10.7554/eLife.52868} {\bibfield  {journal} {\bibinfo  {journal} {eLife}\
  }\textbf {\bibinfo {volume} {9}},\ \bibinfo {pages} {e52868} (\bibinfo {year}
  {2020}{\natexlab{a}})}\BibitemShut {NoStop}%
\bibitem [{\citenamefont {Afanzar}\ \emph {et~al.}(2020)\citenamefont
  {Afanzar}, \citenamefont {Buss}, \citenamefont {Stearns},\ and\ \citenamefont
  {Ferrell}}]{afanzar_nucleus_2020}%
  \BibitemOpen
  \bibfield  {author} {\bibinfo {author} {\bibfnamefont {O.}~\bibnamefont
  {Afanzar}}, \bibinfo {author} {\bibfnamefont {G.~K.}\ \bibnamefont {Buss}},
  \bibinfo {author} {\bibfnamefont {T.}~\bibnamefont {Stearns}}, \ and\
  \bibinfo {author} {\bibfnamefont {J.~E.}\ \bibnamefont {Ferrell}},\ }\href
  {\doibase 10.1101/2020.06.16.153437} {\bibfield  {journal} {\bibinfo
  {journal} {bioRxiv}\ ,\ \bibinfo {pages} {2020.06.16.153437}} (\bibinfo
  {year} {2020})}\BibitemShut {NoStop}%
\bibitem [{\citenamefont {Beta}\ and\ \citenamefont
  {Kruse}(2017)}]{beta_intracellular_2017}%
  \BibitemOpen
  \bibfield  {author} {\bibinfo {author} {\bibfnamefont {C.}~\bibnamefont
  {Beta}}\ and\ \bibinfo {author} {\bibfnamefont {K.}~\bibnamefont {Kruse}},\
  }\href {\doibase 10.1146/annurev-conmatphys-031016-025210} {\bibfield
  {journal} {\bibinfo  {journal} {Annual Review of Condensed Matter Physics}\
  }\textbf {\bibinfo {volume} {8}},\ \bibinfo {pages} {239} (\bibinfo {year}
  {2017})}\BibitemShut {NoStop}%
\bibitem [{\citenamefont {Deneke}\ and\ \citenamefont
  {Talia}(2018)}]{deneke_chemical_2018}%
  \BibitemOpen
  \bibfield  {author} {\bibinfo {author} {\bibfnamefont {V.~E.}\ \bibnamefont
  {Deneke}}\ and\ \bibinfo {author} {\bibfnamefont {S.~D.}\ \bibnamefont
  {Talia}},\ }\href {\doibase 10.1083/jcb.201701158} {\bibfield  {journal}
  {\bibinfo  {journal} {J Cell Biol}\ ,\ \bibinfo {pages} {jcb.201701158}}
  (\bibinfo {year} {2018})}\BibitemShut {NoStop}%
\bibitem [{\citenamefont {Gelens}\ \emph {et~al.}(2014)\citenamefont {Gelens},
  \citenamefont {Anderson},\ and\ \citenamefont
  {Ferrell}}]{gelens_spatial_2014}%
  \BibitemOpen
  \bibfield  {author} {\bibinfo {author} {\bibfnamefont {L.}~\bibnamefont
  {Gelens}}, \bibinfo {author} {\bibfnamefont {G.~A.}\ \bibnamefont
  {Anderson}}, \ and\ \bibinfo {author} {\bibfnamefont {J.~E.}\ \bibnamefont
  {Ferrell}},\ }\href {\doibase 10.1091/mbc.E14-08-1306} {\bibfield  {journal}
  {\bibinfo  {journal} {Molecular Biology of the Cell}\ }\textbf {\bibinfo
  {volume} {25}},\ \bibinfo {pages} {3486} (\bibinfo {year}
  {2014})}\BibitemShut {NoStop}%
\bibitem [{\citenamefont {Aranson}\ and\ \citenamefont
  {Kramer}(2002)}]{aranson_world_2002}%
  \BibitemOpen
  \bibfield  {author} {\bibinfo {author} {\bibfnamefont {I.~S.}\ \bibnamefont
  {Aranson}}\ and\ \bibinfo {author} {\bibfnamefont {L.}~\bibnamefont
  {Kramer}},\ }\href {\doibase 10.1103/RevModPhys.74.99} {\bibfield  {journal}
  {\bibinfo  {journal} {Reviews of Modern Physics}\ }\textbf {\bibinfo {volume}
  {74}},\ \bibinfo {pages} {99} (\bibinfo {year} {2002})}\BibitemShut {NoStop}%
\bibitem [{\citenamefont {{Garc{\'i}a-Morales}}\ and\ \citenamefont
  {Krischer}(2012)}]{garcia-morales_complex_2012}%
  \BibitemOpen
  \bibfield  {author} {\bibinfo {author} {\bibfnamefont {V.}~\bibnamefont
  {{Garc{\'i}a-Morales}}}\ and\ \bibinfo {author} {\bibfnamefont
  {K.}~\bibnamefont {Krischer}},\ }\href {\doibase
  10.1080/00107514.2011.642554} {\bibfield  {journal} {\bibinfo  {journal}
  {Contemporary Physics}\ }\textbf {\bibinfo {volume} {53}},\ \bibinfo {pages}
  {79} (\bibinfo {year} {2012})}\BibitemShut {NoStop}%
\bibitem [{\citenamefont {Stich}\ and\ \citenamefont
  {Mikhailov}(2002)}]{stich_complex_2002}%
  \BibitemOpen
  \bibfield  {author} {\bibinfo {author} {\bibfnamefont {M.}~\bibnamefont
  {Stich}}\ and\ \bibinfo {author} {\bibfnamefont {A.~S.}\ \bibnamefont
  {Mikhailov}},\ }\href {\doibase 10.1524/zpch.2002.216.4.521} {\bibfield
  {journal} {\bibinfo  {journal} {Zeitschrift f\"ur Physikalische Chemie}\
  }\textbf {\bibinfo {volume} {216}},\ \bibinfo {pages} {521} (\bibinfo {year}
  {2002})}\BibitemShut {NoStop}%
\bibitem [{\citenamefont {Stich}\ and\ \citenamefont
  {Mikhailov}(2006)}]{stich_target_2006}%
  \BibitemOpen
  \bibfield  {author} {\bibinfo {author} {\bibfnamefont {M.}~\bibnamefont
  {Stich}}\ and\ \bibinfo {author} {\bibfnamefont {A.~S.}\ \bibnamefont
  {Mikhailov}},\ }\href {\doibase 10.1016/j.physd.2006.01.011} {\bibfield
  {journal} {\bibinfo  {journal} {Physica D: Nonlinear Phenomena}\ }\textbf
  {\bibinfo {volume} {215}},\ \bibinfo {pages} {38} (\bibinfo {year}
  {2006})}\BibitemShut {NoStop}%
\bibitem [{\citenamefont {Kopell}(1981)}]{kopell_target_1981}%
  \BibitemOpen
  \bibfield  {author} {\bibinfo {author} {\bibfnamefont {N.}~\bibnamefont
  {Kopell}},\ }\href {\doibase 10.1016/0196-8858(81)90041-5} {\bibfield
  {journal} {\bibinfo  {journal} {Advances in Applied Mathematics}\ }\textbf
  {\bibinfo {volume} {2}},\ \bibinfo {pages} {389} (\bibinfo {year}
  {1981})}\BibitemShut {NoStop}%
\bibitem [{\citenamefont {Neu}()}]{neu_chemical_1979}%
  \BibitemOpen
  \bibfield  {author} {\bibinfo {author} {\bibfnamefont {J.~C.}\ \bibnamefont
  {Neu}},\ }\href {\doibase http://dx.doi.org/10.1137/0136038} {\bibfield
  {journal} {\bibinfo  {journal} {SIAM Journal on Applied Mathematics;
  Philadelphia}\ }\textbf {\bibinfo {volume} {36}},\ \bibinfo {pages}
  {7}}\BibitemShut {NoStop}%
\bibitem [{\citenamefont {Hagan}()}]{hagan_target_1981}%
  \BibitemOpen
  \bibfield  {author} {\bibinfo {author} {\bibfnamefont {P.~S.}\ \bibnamefont
  {Hagan}},\ }\href {\doibase 10.1016/0196-8858(81)90042-7} {\bibfield
  {journal} {\bibinfo  {journal} {Advances in Applied Mathematics}\ }\textbf
  {\bibinfo {volume} {2}},\ \bibinfo {pages} {400}}\BibitemShut {NoStop}%
\bibitem [{\citenamefont {Kuramoto}()}]{kuramoto_chemical_1984}%
  \BibitemOpen
  \bibfield  {author} {\bibinfo {author} {\bibfnamefont {Y.}~\bibnamefont
  {Kuramoto}},\ }\href@noop {} {\emph {\bibinfo {title} {Chemical
  {{Oscillations}}, {{Waves}}, and {{Turbulence}}}}}\ (\bibinfo  {publisher}
  {{Springer}})\BibitemShut {NoStop}%
\bibitem [{\citenamefont {Tyson}\ and\ \citenamefont
  {Keener}()}]{tyson_singular_1988}%
  \BibitemOpen
  \bibfield  {author} {\bibinfo {author} {\bibfnamefont {J.~J.}\ \bibnamefont
  {Tyson}}\ and\ \bibinfo {author} {\bibfnamefont {J.~P.}\ \bibnamefont
  {Keener}},\ }\href {\doibase 10.1016/0167-2789(88)90062-0} {\bibfield
  {journal} {\bibinfo  {journal} {Physica D: Nonlinear Phenomena}\ }\textbf
  {\bibinfo {volume} {32}},\ \bibinfo {pages} {327}}\BibitemShut {NoStop}%
\bibitem [{\citenamefont {Meron}(1992)}]{meron_pattern_1992}%
  \BibitemOpen
  \bibfield  {author} {\bibinfo {author} {\bibfnamefont {E.}~\bibnamefont
  {Meron}},\ }\href {\doibase 10.1016/0370-1573(92)90098-K} {\bibfield
  {journal} {\bibinfo  {journal} {Physics Reports}\ }\textbf {\bibinfo {volume}
  {218}},\ \bibinfo {pages} {1} (\bibinfo {year} {1992})}\BibitemShut {NoStop}%
\bibitem [{\citenamefont {Casten}\ \emph {et~al.}(1975)\citenamefont {Casten},
  \citenamefont {Cohen},\ and\ \citenamefont
  {Lagerstrom}}]{casten_perturbation_1975}%
  \BibitemOpen
  \bibfield  {author} {\bibinfo {author} {\bibnamefont {Casten}}, \bibinfo
  {author} {\bibnamefont {Cohen}}, \ and\ \bibinfo {author} {\bibnamefont
  {Lagerstrom}},\ }\href@noop {} {\bibfield  {journal} {\bibinfo  {journal}
  {Quarterly of Applied Mathematics}\ }\textbf {\bibinfo {volume} {32}},\
  \bibinfo {pages} {365} (\bibinfo {year} {1975})}\BibitemShut {NoStop}%
\bibitem [{\citenamefont {Keener}(1986)}]{keener_geometrical_1986}%
  \BibitemOpen
  \bibfield  {author} {\bibinfo {author} {\bibfnamefont {J.}~\bibnamefont
  {Keener}},\ }\href {\doibase 10.1137/0146062} {\bibfield  {journal} {\bibinfo
   {journal} {SIAM Journal on Applied Mathematics}\ }\textbf {\bibinfo {volume}
  {46}},\ \bibinfo {pages} {1039} (\bibinfo {year} {1986})}\BibitemShut
  {NoStop}%
\bibitem [{\citenamefont {Keener}\ and\ \citenamefont
  {Sneyd}()}]{keener_mathematical_2009}%
  \BibitemOpen
  \bibfield  {author} {\bibinfo {author} {\bibfnamefont {J.}~\bibnamefont
  {Keener}}\ and\ \bibinfo {author} {\bibfnamefont {J.}~\bibnamefont {Sneyd}},\
  }\href {\doibase 10.1007/978-0-387-75847-3} {\emph {\bibinfo {title}
  {Mathematical {{Physiology}}: {{I}}: {{Cellular Physiology}}}}},\ \bibinfo
  {edition} {2nd}\ ed.,\ Interdisciplinary {{Applied Mathematics}}\ (\bibinfo
  {publisher} {{Springer-Verlag}})\BibitemShut {NoStop}%
\bibitem [{\citenamefont {{Van der
  Pol}}(1926)}]{van_der_pol_relaxation-oscillations_1926}%
  \BibitemOpen
  \bibfield  {author} {\bibinfo {author} {\bibfnamefont {B.}~\bibnamefont {{Van
  der Pol}}},\ }\href {\doibase 10.1080/14786442608564127} {\bibfield
  {journal} {\bibinfo  {journal} {The London, Edinburgh, and Dublin
  Philosophical Magazine and Journal of Science}\ }\textbf {\bibinfo {volume}
  {2}},\ \bibinfo {pages} {978} (\bibinfo {year} {1926})}\BibitemShut {NoStop}%
\bibitem [{\citenamefont {FitzHugh}(1961)}]{fitzhugh_impulses_1961}%
  \BibitemOpen
  \bibfield  {author} {\bibinfo {author} {\bibfnamefont {R.}~\bibnamefont
  {FitzHugh}},\ }\href@noop {} {\bibfield  {journal} {\bibinfo  {journal}
  {Biophysical Journal}\ }\textbf {\bibinfo {volume} {1}},\ \bibinfo {pages}
  {445} (\bibinfo {year} {1961})}\BibitemShut {NoStop}%
\bibitem [{\citenamefont {Nagumo}\ \emph {et~al.}(1962)\citenamefont {Nagumo},
  \citenamefont {Arimoto},\ and\ \citenamefont
  {Yoshizawa}}]{nagumo_active_1962}%
  \BibitemOpen
  \bibfield  {author} {\bibinfo {author} {\bibfnamefont {J.}~\bibnamefont
  {Nagumo}}, \bibinfo {author} {\bibfnamefont {S.}~\bibnamefont {Arimoto}}, \
  and\ \bibinfo {author} {\bibfnamefont {S.}~\bibnamefont {Yoshizawa}},\ }\href
  {\doibase 10.1109/JRPROC.1962.288235} {\bibfield  {journal} {\bibinfo
  {journal} {Proceedings of the Institute of Radio Engineers}\ }\textbf
  {\bibinfo {volume} {50}},\ \bibinfo {pages} {2061} (\bibinfo {year}
  {1962})}\BibitemShut {NoStop}%
\bibitem [{\citenamefont {Winfree}(1967)}]{winfree_biological_1967}%
  \BibitemOpen
  \bibfield  {author} {\bibinfo {author} {\bibfnamefont {A.~T.}\ \bibnamefont
  {Winfree}},\ }\href {\doibase 10.1016/0022-5193(67)90051-3} {\bibfield
  {journal} {\bibinfo  {journal} {Journal of Theoretical Biology}\ }\textbf
  {\bibinfo {volume} {16}},\ \bibinfo {pages} {15} (\bibinfo {year}
  {1967})}\BibitemShut {NoStop}%
\bibitem [{\citenamefont {Nakao}()}]{nakao_phase_2016}%
  \BibitemOpen
  \bibfield  {author} {\bibinfo {author} {\bibfnamefont {H.}~\bibnamefont
  {Nakao}},\ }\href {\doibase 10.1080/00107514.2015.1094987} {\bibfield
  {journal} {\bibinfo  {journal} {Contemporary Physics}\ }\textbf {\bibinfo
  {volume} {57}},\ \bibinfo {pages} {188}}\BibitemShut {NoStop}%
\bibitem [{\citenamefont {Ermentrout}\ and\ \citenamefont
  {Terman}()}]{ermentrout_mathematical_2010}%
  \BibitemOpen
  \bibfield  {author} {\bibinfo {author} {\bibfnamefont {G.~B.}\ \bibnamefont
  {Ermentrout}}\ and\ \bibinfo {author} {\bibfnamefont {D.~H.}\ \bibnamefont
  {Terman}},\ }\href@noop {} {\emph {\bibinfo {title} {Mathematical
  {{Foundations}} of {{Neuroscience}}}}},\ \bibinfo {series} {Interdisciplinary
  {{Applied Mathematics}}}, Vol.~\bibinfo {volume} {35}\ (\bibinfo  {publisher}
  {{Springer}})\BibitemShut {NoStop}%
\bibitem [{\citenamefont {Kotani}\ \emph {et~al.}()\citenamefont {Kotani},
  \citenamefont {Yamaguchi}, \citenamefont {Ogawa}, \citenamefont {Jimbo},
  \citenamefont {Nakao},\ and\ \citenamefont
  {Ermentrout}}]{kotani_adjoint_2012}%
  \BibitemOpen
  \bibfield  {author} {\bibinfo {author} {\bibfnamefont {K.}~\bibnamefont
  {Kotani}}, \bibinfo {author} {\bibfnamefont {I.}~\bibnamefont {Yamaguchi}},
  \bibinfo {author} {\bibfnamefont {Y.}~\bibnamefont {Ogawa}}, \bibinfo
  {author} {\bibfnamefont {Y.}~\bibnamefont {Jimbo}}, \bibinfo {author}
  {\bibfnamefont {H.}~\bibnamefont {Nakao}}, \ and\ \bibinfo {author}
  {\bibfnamefont {G.~B.}\ \bibnamefont {Ermentrout}},\ }\href {\doibase
  10.1103/PhysRevLett.109.044101} {\bibfield  {journal} {\bibinfo  {journal}
  {Physical Review Letters}\ }\textbf {\bibinfo {volume} {109}},\ \bibinfo
  {pages} {044101}}\BibitemShut {NoStop}%
\bibitem [{\citenamefont {Novičenko}\ and\ \citenamefont
  {Pyragas}()}]{novicenko_phase_2012}%
  \BibitemOpen
  \bibfield  {author} {\bibinfo {author} {\bibfnamefont {V.}~\bibnamefont
  {Novičenko}}\ and\ \bibinfo {author} {\bibfnamefont {K.}~\bibnamefont
  {Pyragas}},\ }\href {\doibase 10.1016/j.physd.2012.03.001} {\bibfield
  {journal} {\bibinfo  {journal} {Physica D: Nonlinear Phenomena}\ }\textbf
  {\bibinfo {volume} {241}},\ \bibinfo {pages} {1090}}\BibitemShut {NoStop}%
\bibitem [{\citenamefont {{de Alcantara Bonfim}}\ and\ \citenamefont
  {Griffiths}(2005)}]{de_alcantara_bonfim_exact_2005}%
  \BibitemOpen
  \bibfield  {author} {\bibinfo {author} {\bibfnamefont {O.~F.}\ \bibnamefont
  {{de Alcantara Bonfim}}}\ and\ \bibinfo {author} {\bibfnamefont {D.~J.}\
  \bibnamefont {Griffiths}},\ }\href {\doibase 10.1119/1.2140771} {\bibfield
  {journal} {\bibinfo  {journal} {American Journal of Physics}\ }\textbf
  {\bibinfo {volume} {74}},\ \bibinfo {pages} {43} (\bibinfo {year}
  {2005})}\BibitemShut {NoStop}%
\bibitem [{\citenamefont {Luther}(1906)}]{luther_raumliche_1906}%
  \BibitemOpen
  \bibfield  {author} {\bibinfo {author} {\bibfnamefont {R.}~\bibnamefont
  {Luther}},\ }\href {\doibase 10.1002/bbpc.19060123208} {\bibfield  {journal}
  {\bibinfo  {journal} {Zeitschrift f\"ur Elektrochemie und angewandte
  physikalische Chemie}\ }\textbf {\bibinfo {volume} {12}},\ \bibinfo {pages}
  {596} (\bibinfo {year} {1906})}\BibitemShut {NoStop}%
\bibitem [{\citenamefont {Arnold}\ \emph {et~al.}(1987)\citenamefont {Arnold},
  \citenamefont {Showalter},\ and\ \citenamefont
  {Tyson}}]{arnold_propagation_1987}%
  \BibitemOpen
  \bibfield  {author} {\bibinfo {author} {\bibfnamefont {R.}~\bibnamefont
  {Arnold}}, \bibinfo {author} {\bibfnamefont {K.}~\bibnamefont {Showalter}}, \
  and\ \bibinfo {author} {\bibfnamefont {J.~J.}\ \bibnamefont {Tyson}},\ }\href
  {\doibase 10.1021/ed064p740} {\bibfield  {journal} {\bibinfo  {journal}
  {Journal of Chemical Education}\ }\textbf {\bibinfo {volume} {64}},\ \bibinfo
  {pages} {740} (\bibinfo {year} {1987})}\BibitemShut {NoStop}%
\bibitem [{\citenamefont {Showalter}\ and\ \citenamefont
  {Tyson}(1987)}]{showalter_luthers_1987}%
  \BibitemOpen
  \bibfield  {author} {\bibinfo {author} {\bibfnamefont {K.}~\bibnamefont
  {Showalter}}\ and\ \bibinfo {author} {\bibfnamefont {J.~J.}\ \bibnamefont
  {Tyson}},\ }\href {\doibase 10.1021/ed064p742} {\bibfield  {journal}
  {\bibinfo  {journal} {Journal of Chemical Education}\ }\textbf {\bibinfo
  {volume} {64}},\ \bibinfo {pages} {742} (\bibinfo {year} {1987})}\BibitemShut
  {NoStop}%
\bibitem [{\citenamefont {Kotani}\ \emph {et~al.}(2020)\citenamefont {Kotani},
  \citenamefont {Ogawa}, \citenamefont {Shirasaka}, \citenamefont {Akao},
  \citenamefont {Jimbo},\ and\ \citenamefont {Nakao}}]{kotani_nonlinear_2020}%
  \BibitemOpen
  \bibfield  {author} {\bibinfo {author} {\bibfnamefont {K.}~\bibnamefont
  {Kotani}}, \bibinfo {author} {\bibfnamefont {Y.}~\bibnamefont {Ogawa}},
  \bibinfo {author} {\bibfnamefont {S.}~\bibnamefont {Shirasaka}}, \bibinfo
  {author} {\bibfnamefont {A.}~\bibnamefont {Akao}}, \bibinfo {author}
  {\bibfnamefont {Y.}~\bibnamefont {Jimbo}}, \ and\ \bibinfo {author}
  {\bibfnamefont {H.}~\bibnamefont {Nakao}},\ }\href {\doibase
  10.1103/PhysRevResearch.2.033106} {\bibfield  {journal} {\bibinfo  {journal}
  {Physical Review Research}\ }\textbf {\bibinfo {volume} {2}},\ \bibinfo
  {pages} {033106} (\bibinfo {year} {2020})}\BibitemShut {NoStop}%
\bibitem [{\citenamefont {Mahara}\ \emph {et~al.}(2000)\citenamefont {Mahara},
  \citenamefont {Saito}, \citenamefont {Amagishi}, \citenamefont {Nagashima},\
  and\ \citenamefont {Yamaguchi}}]{mahara_ring-shaped_2000}%
  \BibitemOpen
  \bibfield  {author} {\bibinfo {author} {\bibfnamefont {H.}~\bibnamefont
  {Mahara}}, \bibinfo {author} {\bibfnamefont {T.}~\bibnamefont {Saito}},
  \bibinfo {author} {\bibfnamefont {Y.}~\bibnamefont {Amagishi}}, \bibinfo
  {author} {\bibfnamefont {H.}~\bibnamefont {Nagashima}}, \ and\ \bibinfo
  {author} {\bibfnamefont {T.}~\bibnamefont {Yamaguchi}},\ }\href {\doibase
  10.1143/JPSJ.69.3552} {\bibfield  {journal} {\bibinfo  {journal} {Journal of
  the Physical Society of Japan}\ }\textbf {\bibinfo {volume} {69}},\ \bibinfo
  {pages} {3552} (\bibinfo {year} {2000})}\BibitemShut {NoStop}%
\bibitem [{\citenamefont {Nolet}\ \emph
  {et~al.}(2020{\natexlab{b}})\citenamefont {Nolet}, \citenamefont {Rombouts},\
  and\ \citenamefont {Gelens}}]{nolet_synchronization_2020}%
  \BibitemOpen
  \bibfield  {author} {\bibinfo {author} {\bibfnamefont {F.~E.}\ \bibnamefont
  {Nolet}}, \bibinfo {author} {\bibfnamefont {J.}~\bibnamefont {Rombouts}}, \
  and\ \bibinfo {author} {\bibfnamefont {L.}~\bibnamefont {Gelens}},\ }\href
  {\doibase 10.1063/5.0002251} {\bibfield  {journal} {\bibinfo  {journal}
  {Chaos: An Interdisciplinary Journal of Nonlinear Science}\ }\textbf
  {\bibinfo {volume} {30}},\ \bibinfo {pages} {053139} (\bibinfo {year}
  {2020}{\natexlab{b}})}\BibitemShut {NoStop}%
\bibitem [{\citenamefont {T{\"o}njes}\ and\ \citenamefont
  {Blasius}(2009)}]{tonjes_perturbation_2009}%
  \BibitemOpen
  \bibfield  {author} {\bibinfo {author} {\bibfnamefont {R.}~\bibnamefont
  {T{\"o}njes}}\ and\ \bibinfo {author} {\bibfnamefont {B.}~\bibnamefont
  {Blasius}},\ }\href {\doibase 10.1103/PhysRevE.79.016112} {\bibfield
  {journal} {\bibinfo  {journal} {Physical Review E}\ }\textbf {\bibinfo
  {volume} {79}},\ \bibinfo {pages} {016112} (\bibinfo {year}
  {2009})}\BibitemShut {NoStop}%
\end{thebibliography}

%

\end{document}